\documentstyle[11pt,epsf]{article}
\newcommand{\gsim}{\mathrel{\raise.3ex\hbox{$>$\kern-.75em\lower1ex
\hbox{$\sim$}}}}
\newcommand{\lesssim}{\mathrel{\raise.3ex\hbox{$<$\kern-.75em
\lower1ex\hbox{$\sim$}}}}

\newcommand{\be}{\begin{equation}}
\newcommand{\ee}{\end{equation}}
\newcommand{\bea}{\begin{eqnarray}}
\newcommand{\eea}{\end{eqnarray}}

\newcommand{\Dx}{d \Pi_\chi}
\newcommand{\DX}{d \Pi_{\bar\chi}}
\newcommand{\Db}{d \Pi_b}
\newcommand{\DB}{d \Pi_{\bar b}}
\newcommand{\Dd}{d \Pi_d}
\newcommand{\DD}{d \Pi_{\bar d}}
\newcommand{\Ds}{d \Pi_\sigma}
\newcommand{\DS}{d \Pi_{\bar\sigma}}
\newcommand{\Do}{d \Pi_1}
\newcommand{\Dt}{d \Pi_2}
\newcommand{\DT}{d \Pi_3}
\newcommand{\Df}{d \Pi_4}
\newcommand{\DF}{d \Pi_5}
\newcommand{\Dk}{d \Pi_*}
\newcommand{\Da}{d \Pi_{\alpha_1}}
\newcommand{\DA}{d \Pi_{\alpha_2}}
\newcommand{\Dc}{d \Pi_{\beta_1}}
\newcommand{\DC}{d \Pi_{\beta_2}}

\begin{document}

\begin{titlepage}

\begin{flushright}
CINVESTAV-FIS-03/96
\end{flushright}

\begin{center}{\Large\bf Particle-Antiparticle Asymmetry Due to\\[0.5ex]
  Non-Renormalizable Effective Interactions}\\[2cm]
\end{center}

\begin{center}
  {\large\bf Antonio O.\ Bouzas,}$^{\mbox{\small a~1}}$
  {\large\bf Wai Yan Cheng,}$^{\mbox{\small b~2}}$\\ 
  {\large\bf Graciela Gelmini.}$^{\mbox{\small b~3}}$\\[2ex]
  $^{\mbox{\small a}}$Departamento de F\'{\i}sica, Centro de
  Investigaci\'on y Estudios Avanzados, Instituto Polit\'ecnico
  Nacional, Apartado Postal 14--740, \mbox{07000 M\'exico D.\ F.}\\[2ex]
  $^{\mbox{\small b}}$Department of Physics,
  University of California, Los Angeles,\\ California 90024-1547
\end{center}
\vspace{2cm}

\begin{abstract}
We consider a model for generating a particle-antiparticle asymmetry
through out-of-equilibrium decays of a massive particle due to
non-renormalizable, effective interactions.  
\end{abstract}

\footnotetext[1]{E-mail: abouzas@fis.cinvestav.mx} 
\footnotetext[2]{E-mail: wcheng@physics.ucla.edu}
\footnotetext[3]{E-mail: gelmini@sting.physics.ucla.edu}

\end{titlepage}

\section{Introduction}
\setcounter{equation}{0}

We study in this paper the generation of a fermion number $(F)$
asymmetry in the  
early universe due to $F$-violations in non-renormalizable effective 
interactions. We are motivated  by the possibility of generating the 
cosmological baryon number asymmetry through violation of 
global symmetries, such as global baryon or lepton numbers, by quantum
gravity \cite{geho}.
However, we will not restrict ourselves to this case. We consider the 
generic case of effective interactions due to physics at a scale 
$\Lambda \leq M_P$.

We have constructed a minimal toy model that, we expect, shows the main 
features of $F$-generation in generic models with
$F$-violation only in non-renormalizable interactions.
The toy model consists of fermions $b$  and $d$, and two scalars $\chi$ and 
$\sigma$.
The $F$-asymmetry is generated through out-of-equilibrium decays of 
the massive $\chi$.
All other fields are taken to be massless for simplicity.
It is assumed that other non-specified interactions maintain 
equilibrium distributions for all particles except $\chi$.

The different scenarios of $F$-generation are classified by ranges of 
values of the following constants. 
\begin{itemize}
\item $K\equiv(\Gamma_\chi/H)_{T=M}$, the ``effectiveness of decay'' 
parameter, and
\item $Kr_0 \simeq \Gamma_\chi^{NR}/H$, the
``effectiveness of the  non-renormalizable interactions,''
where $r_0 \simeq \Gamma_\chi^{NR}/\Gamma_\chi$.
\end{itemize}
 $K$ is given by the ratio of the decay rate $\Gamma_\chi$ of $\chi$ to
the Hubble constant at the moment when the $\chi$ bosons are becoming
non-relativistic, approximately when the temperature is equal to their
mass, $M$.  $r_0$ is the ratio of the $\chi$ decay width due to 
non-renormalizable decays through $F$-violating interactions, 
$\Gamma_\chi^{NR}$, 
to the total width, $\Gamma_\chi$. For the precise definition of $K$
and $r_0$ see (\ref{eq:kexact}) and (\ref{eq:r0}). 

The total $\chi$ decay
width, $\Gamma_\chi$, is dominated by renormalizable interactions, whose 
coupling we call $g_1$. Thus $\Gamma_\chi(T \leq M) \simeq g_1^2 M/8
\pi$ (see (\ref{eq:dominated})). We call $g_2$ the coupling constant of the
non-renormalizable $\chi$  interactions that provide $F$-violating $\chi$
decay, namely $\Gamma_\chi^{NR}(T\leq M) \simeq g_2^2 M^3/\Lambda^2$.
Taking $g_*$, the effective number of relativistic degrees of freedom
appearing in the 
Hubble expansion rate \cite{kotu}, $H \simeq \sqrt{g_*}T^2/M_P$, 
from now on to be a reasonable number $g_* \simeq 10^2$, 
we get for $K$ and $r_0$ (see  (\ref{eq:kapprox}) and (\ref{eq:r0approx})), 
\be \label{eq:firsteq}
K\simeq 10^{-2} g_1^2(M_p/M),~~~~~~~
r_0 \simeq (g_2 M/g_1 \Lambda)^2~~.
\ee

The other independent parameters that completely define the resulting 
$F$-asymmetry are, 
\begin{itemize}
\item[$\ast$] $g_1$, the coupling of the renormalizable interactions 
that dominate the $\chi$ decay rate, and
\item[$\ast$] $\eta$ and $\xi$, two parameters related to $CP$-violation in
  decays and annihilation processes, respectively (see below).
\end{itemize}
We have chosen to keep these last three parameters constant at reasonable but 
arbitrary values, $g_1 = 10^{-1} $ and $\eta = \xi = 5 ~ 10^{-4}$, in most 
of the cases presented below, to show the effects of the different ranges 
of the main parameters $K$ and $Kr_0$.
In most cases the effects of changing $g_1$, $\eta$ and $\xi$ can be 
easily understood.

        In order to be consistent with our Lagrangian, 
in which the effects of 
the physics at scale $\Lambda$ only remain in effective interactions, we 
have to consider energies below $\Lambda$, thus $T \leq \Lambda$. Since  
also $M \lesssim \Lambda$, from (\ref{eq:firsteq}) we deduce that small values of $K$, 
 $K \ll 1$, can only be obtained for $\Lambda$ not much lower than $M_P$, so 
that $M$ $\simeq \Lambda \simeq M_P$ 
and $K$ becomes not much larger than $10^{-2}g_1^2$.
Given that $r_0 \leq 1$ always, in the case $K \ll 1$
we necessarily have $Kr_0 \ll 1$. Values of $K$ not 
much smaller than 1 can be obtained 
for any value of $\Lambda$.  As we 
will see, the main remaining ranges for which qualitatively different 
$F$-asymmetry generation patterns are obtained are
\begin{itemize}
\item[-] $K r_0 \lesssim 10^{-2}$,
\item[-] $10^{-2} \lesssim K r_0 \lesssim 10^7$,
\item[-] $10^7 \lesssim K $.
\end{itemize}

While the usual, renormalizable scenarios of $B$-violation can be 
classified with the sole parameter $K$ \cite{kowo,frol,hako}, that
mainly regulates the  
departure of the $\chi$-abundance, $Y_\chi\equiv n_\chi/n_\gamma$ 
($n$ stands for number density), with respect 
to its equilibrium abundance $Y_\chi^{eq}$, it is obvious that here $Kr_0$ is 
also important.
This is so because only non-renormalizable interactions generate a net
$F$-abundance, $Y_F$, whose ``effectiveness'' is given by 
$Kr_0 \simeq \Gamma_\chi^{NR}/H$, as will become clear below. 

\section{The Model}
\setcounter{equation}{0}

The schematic model we consider in this paper belongs in the so-called
standard scenario of out-of-equilibrium decays.  As is well known, in
such scenario three elements are necessary in order to generate
dynamically an $F$-asymmetry \cite{sakh,wein}.  Namely, an
 $F$-violating interaction, violation of $C$ and $CP$ symmetries, and a
departure from thermal equilibrium.  The last ingredient is provided
by the expansion of the Universe, which we assume to be described by
the Friedmann-Robertson-Walker cosmology, in the context of the hot
big bang model \cite{kotu,kotu2,oliv}.  The other two are discussed
below. 

\subsection{Lagrangian}

We include two different fermion species $b$ and $d$ to simplify 
calculations by avoiding the presence of cumbersome interference terms 
that are irrelevant for the classification of different 
$F$-generation scenarios, which is our main objective.
Two different bosons $\chi$ and $\sigma$, with wisely chosen $F$-number, 
give rise to simple $F$-violating dimension five operators, containing 
$\chi\sigma$ and two fermions, while allowing for $F$-conserving 
renormalizable Yukawa couplings of $\chi$.
This is a feature we want to preserve, namely, we expect that in 
generic models the decaying boson $\chi$ will have both renormalizable 
as well as non-renormalizable decays.
Renormalizable Yukawa couplings of the second boson, $\sigma$, would 
only unnecessarily complicate the model, thus we do not include them.
Therefore, trilinear couplings $\chi\chi\sigma$ are included in the scalar 
potential to define a non-zero $F$-number for $\sigma$, $F_\sigma = 
-2 F_\chi$, while $F_\chi$ is defined in the Majorana type Yukawa 
coupling of $\chi$, $\bar{b^c} d \chi^\dagger$, to be $F_\chi = F_d + F_b$.
By arbitrarily choosing $F_d = F_b = 1$ one obtains $F_\chi = 2$ and 
$F_\sigma = -4$ from the above couplings.

The complete Lagrangian of our toy model is, then,
\be   \label{eq:lagrangian}
{\cal L}  =   g_1  (\bar{b^c}   d   \chi^\dagger   + \mbox{h.c.})     +
\frac{g_2}{\Lambda} (\bar{b^c}   d \chi^\dagger  \sigma   +  \mbox{h.c.})   +
V(\chi,\sigma)~~, 
\ee
with
\be
V(\chi,\sigma)  = - M^2 \chi^\dagger\chi + g_3 (\chi\chi\sigma + 
\chi^\dagger\chi^\dagger\sigma^\dagger) 
 + g_4 \chi^\dagger\chi\sigma^\dagger \sigma~~.
\ee
Several comments are in order.
We have assumed the existence of only one non-renormalizable 
$F$-violating term.
It could be argued that, for example, quantum gravity could generate 
also $F$-violating renormalizable terms.
Even though no strong argument can be given against this possibility, its 
effect would be so severe that any approximate conservation of a global 
number would be invalidated.
Inclusion of other $F$-violating terms with dimensions $\geq 5$ 
would enormously complicate the model, and we would like to keep it as 
simple as possible.  At any rate, dimension five operators will be
dominant in most models. (Operators of dim-5 and 6 have been
considered in the framework of SUSY GUTS in \cite{frtu,kora}.)

Even after these considerations the term
\be
{g_2^\prime\over\Lambda} (\bar{b^c} d\chi\sigma + h.c.)
\ee
should legitimately be included in (\ref{eq:lagrangian}).
However, the effects of this term would be the same as the $g_2$ term in 
(\ref{eq:lagrangian}).
In general $g_2$ and $g_2^\prime$ will not be equal.
If both, the $g_2$ and $g_2^\prime$ terms, are of the same order 
of magnitude, their combined effect would be of the same order as that 
of just one of them.
If, instead, one of them dominates, the other will again not change the 
order of magnitude of their combined effect.
Thus, without loss of generality we assume $g_2^\prime \ll g_2$ and neglect 
the $g_2^\prime$ term.

We will consider the masses of $b$, $d$, and $\sigma$ to be negligibly
small with respect to the mass of $\chi$, $M$.

In spite of its simplicity, the model possesses several reaction channels,
contributing to $\chi$-decay
and inverse-decay, $\chi\bar\chi$ annihilation, $2 \leftrightarrow 2$
``point"-scatterings, due to contact interactions of the incoming and
outgoing particles, and $2 \leftrightarrow 2$ and $2 \leftrightarrow
 3$ scatterings.  Two-to-three scatterings, in particular, give rise
to a rather large number of diagrams.

The $F$-asymmetry is generated by decays and annihilations of $\chi$
particles, while the other processes, inverse-decays and scatterings,
partially or completely erase any asymmetry either of dynamical origin
or due to an asymmetric initial condition.  Decays and inverse-decays
are the only relevant processes at small $K$, $K \lesssim 1$.  As $K$
becomes progressively larger than one, other terms in the evolution
equations become important.  Point scatterings, together with
inverse-decays, are the most efficient damping processes, while
$2\leftrightarrow 3$ scatterings are relevant only for extremely large
values of $K$, $Kr_0\gsim 10^7$ with our choice of values of $g_1$,
$\eta$ and $\xi$, simply because only then they dominate over inverse
decays before all interactions other than decays go out of
equilibrium.

\subsection{$CP$ violation}
\label{sec:cpviolation}

The interaction lagrangian (\ref{eq:lagrangian}) does not give rise to
a violation of $CP$ symmetry.  In order to model $CP$ violations, we
should introduce more fields and couplings into (\ref{eq:lagrangian}).
Since we want to keep our model as simple as possible, we shall
instead resort to an explicit parametrization of the squared
amplitudes for various processes, that respects $CPT$ symmetry and
unitarity but violates $CP$ \cite{kowo,frol}.

The branching ratios for $F$-violating decays of $\chi$ and $\bar
\chi$ particles are defined as,
\be
r\equiv {\Gamma (\chi \rightarrow bd\sigma) \over \Gamma_\chi},~~ 
\bar r \equiv {\Gamma(\bar\chi\rightarrow \bar b \bar d 
\bar\sigma) \over \Gamma_\chi}~. 
\ee
Notice that $\Gamma_{\bar\chi} = \Gamma_\chi$ due to $CPT$ invariance.
When  $CP$ is violated  $r$ and $\bar r$ are different. We
introduce the $CP$ violation parameter $\eta$, 
\be \label{eq:eta}
\eta r_0 \equiv {1\over 2}(\bar r- r)  
\ee
where
\be \label{eq:r0}
r_0 \equiv {1\over 2} (r + \bar r)~,
\ee
to describe $CP$ in decays and inverse decays.

Besides $\eta$, we have to consider the possibility of $CP$ violation
in scatterings.  There is no $CP$ violation in ``point"-scatterings
({\em i.e.}, $|\chi\bar\sigma\rangle\rightarrow |bd\rangle$, and
crossed channels) and most of the $2 \leftrightarrow 3$ scatterings
because of unitarity, as explained in appendix B.  However, $CP$
violation is possible in $|b \bar b \rangle \rightarrow |d \bar d
\sigma \rangle$ and its back-process, as well as in $\chi\bar{\chi}$
annihilations.  Moreover, the $CP$-violations in both processes are
related to each other (see appendix B) so that we only need to define
a single parameter $\xi$ as,
\be \label{eq:xi}
 \xi \sigma (\chi \bar\chi \rightarrow b\bar b\sigma)
\equiv {1 \over 2}
\left( \sigma (\chi\bar\chi \rightarrow b\bar b\bar\sigma) - \sigma
(\chi\bar\chi \rightarrow b\bar b\sigma)\right)~~.  
\ee
 Therefore, unitarity relations
leave only two independent $CP$-violation parameters in our model, $\eta$
and $\xi$, that we assume to be independent of $x$. 

\section{Boltzmann Equations}
\setcounter{equation}{0}
\label{sec:boleq}
        We assume that $b,d,\sigma$ and their antiparticles
are kept in kinetic and chemical equilibrium with other particles 
through unspecified fast reactions.
Thus we take the chemical potentials of each of these pairs of 
particle and antiparticle as always equal and opposite, $\mu_b = - 
\mu_{\bar b}$, $\mu_d = - \mu_{\bar d}$ and $\mu_\sigma = - 
\mu_{\bar\sigma}$.
Moreover, we make the simplification of considering $\mu_b = \mu_d$,
in view of the symmetry of the model under interchange of $b$ and $d$.
We further assume that only the interactions in (\ref{eq:lagrangian})
are responsible  
for the evolution of these chemical potentials in the ranges of 
temperature where these interactions are dominant.
Thus, we have, 
\be
f_b(p) = f_d(p) = e^{-(E-\mu_b) / T}, ~~ f_\sigma (p) = 
e^{-(E-\mu_\sigma) / T}~.
\ee

In the range of temperatures of interest for global-charge generation,
the $\chi$ and $\bar\chi$ particles are going out of thermal
equilibrium. This provides the out-of-equilibrium element needed for
the generation of a particle asymmetry \cite{kotu,kotu2,oliv}. Thus,
only for large $T$, $T\gg M$, we can assume that $f_\chi(p)$ will be
close to the thermal equilibrium distribution $f_\chi^{eq} (p) =
e^{-E_\chi / T}$.

It is customary ({\em e.\frenchspacing g.,\/} \cite{kowo}) 
to scale out the effect
of the expansion of the universe by considering ratios of the number
densities $n_i$ to the number density of photons, $Y_i =
 n_i/n_\gamma$.\footnote{We could have equally well used the entropy
density $s = g_{*s}n_\gamma$ instead of $n_\gamma$, since we
consider the effective number of relativistic degrees of freedom
 $g_{*s}$ to be constant over the range of temperatures of interest
to us.} Because the interactions in (\ref{eq:lagrangian}) insure
that
\be \label{eq:insured}
{d\over dt} \left( Y_\chi - Y_{\bar\chi}\right )= - ~{d\over dt}
\left ( Y_b - Y_{\bar b}\right )~,
\ee
as it will become clear below, we will need only to solve for
$Y_\chi$, $Y_{\bar\chi}$, $Y_\sigma$ and $Y_{\bar\sigma}$.
Actually, (\ref{eq:insured}) implies that the net $F$-number abundance
$Y_F$ is just 
\be \label{eq:yfs}
Y_F = 4 \left ( Y_{\bar\sigma} - Y_\sigma\right )
\simeq \left ( Y_{\bar\sigma} - Y_\sigma\right )~.
\ee
Since we are only interested in order-of-magnitude estimates, we will drop 
the factor of four in (\ref{eq:yfs}) from now on (which is equivalent
to effectively  
taking $-F_{\bar\sigma}= F_\sigma=1$).
Thus, we will write below three coupled equations for $Y_F$, and $Y_-$, $Y_+$,
defined as
\be
Y_\pm \equiv \left ( Y_\chi \pm Y_{\bar\chi}\right )/2~.
\ee
This choice of densities is convenient because, as we will see, 
the contribution of $Y_-$ to the evolution of $Y_+$ and $Y_F$ is small,
so that we will drop all terms containing $Y_-$  and 
solve only two coupled equations.

Let us mention some useful relations between the scaled number densities
$Y_i$ and the chemical potentials $\mu_i$.
Because, 
\[
n_\sigma = \frac{1}{(2\pi)^{3}} \int \!\!d^3p~e^{-E/T+\mu_\sigma/T} =
\frac{n_\gamma}{2}e^{+\mu_\sigma/T} 
\]
we have,
\be
 Y_F \simeq Y_{\bar\sigma} - Y_\sigma 
=(e^{-\mu_\sigma/T} - e^{\mu_\sigma/T})/2.
\ee
Since we are looking to produce small asymmetries, $|Y_F| \simeq 
 O(10^{-10})$, it is a good approximation to consider $\mu_b, \mu_\sigma 
\ll T$. Expanding the exponentials we get  $Y_F \simeq - \mu_\sigma/T$ and,
thus,
\be \label{eq:chem1}
e^{\pm\mu_\sigma/T} \simeq 1 \mp Y_F~.
\ee
Similarly, we obtain $e^{\pm \mu_b/T} = 1 \mp [(Y_b-Y_{\bar b})/2]$.
Moreover, from (\ref{eq:insured}) and the initial conditions (at high
enough $T$) $Y_- 
=  0$, $Y_b - Y_{\bar b} = 0$ we obtain  $Y_- =  Y_b - Y_{\bar b}$ and, 
therefore,
\be \label{eq:chem2}
e^{\pm \mu_b/T} \simeq 1 \mp Y_-~.
\ee

\subsection{Evolution Equation for $\chi$-Number Density}

The processes that contribute to the generation of a net number of
$\sigma$ and to its damping are $\chi$ decays $(D)$ and inverse 
decays $(ID)$, both renormalizable $(R)$ and non-renormalizable $(NR)$, 
``point"-scattering $(PS)$ (namely scatterings  due to contact 
interactions of the incoming and outgoing particles, such as
$ \chi \bar\sigma \rightarrow bd$), $\bar\chi \chi$ 
renormalizable annihilations $(RA)$ and non-renormalizable ones ($NRA$, 
such as $\chi\bar\chi \rightarrow b\bar b\sigma$) and their crossed 
channels ($NRCC$, such as $\chi\bar\chi\bar\sigma \rightarrow b\bar b$,
etc). The Boltzmann equation for $n_\chi$ is of the form \cite{kowo,frol}
\be \label{eq:boltzmann}
{dn_\chi\over dt} + 3n_\chi H = - \sum_\alpha \Theta_\alpha
\ee
where $\Theta_\alpha$ are the collision terms due to the processes 
just listed, so that $\alpha$ stands for $\alpha = D,~ID,~ PS,~ RA,$
$~NRA~, NRCC$. The full set of equations is given in  appendix A.

We will now show how to deal with the $\Theta_\alpha$ terms by
considering a few of them in detail \cite{kotu,kowo,frol}.
For example, $\Theta_{D+ID}$ contains the term
\bea
\Theta (\chi \leftrightarrow b_1 d_2 \sigma)
& = &\int \Dx \Db \Dd \Ds
\left [ f_\chi(p_\chi)|M(\chi \rightarrow b d \sigma)|^2 \right.
\nonumber\\
&&\left. \hspace{.2in} - f_b(p_b)f_d(p_d)f_\sigma(p_\sigma)
|M(b d \sigma \rightarrow \chi)|^2\right] ,
\eea
where $M(i\rightarrow j)$ is the Lorentz invariant amplitude for the process 
 $|i \rangle$ going to $|j \rangle$, and $d\Pi_A \equiv 
{(d^4 p_A / (2\pi)^3)}\delta(p_A^2-m_A^2)$). By $CPT$ invariance, 
we can replace $|M(bd\sigma\rightarrow\chi)|^2$ by 
 $|M(\bar\chi\rightarrow\bar b\bar d \bar\sigma)|^2$, and
momentum  conservation  implies 
$f_b(p_1)f_d(p_2)f_\sigma(p_\sigma) = \exp (2\mu_b/T)$ $ \exp (\mu_\sigma/T) f_\chi^{eq}(p_\chi)$, therefore
\bea 
\Theta (\chi \leftrightarrow b_1 d_2 \sigma)
& = & \int \Dx \Db \Dd \Ds
\left [f_\chi(p_\chi)|M(\chi \rightarrow b d \sigma)|^2
\right. \nonumber\\
&& \left. -e^{2\mu_b/T} e^{\mu_\sigma/T}f_\chi^{eq}(p_\chi)
|M(\bar\chi\rightarrow \bar b \bar d 
\bar\sigma)|^2\right ] \nonumber\\
& = & \left [ n_\chi r-(1-2Y_-)(1-Y_F)n_\chi^{eq} \bar r \right ]
\langle\Gamma_\chi\rangle~~, \label{eq:thetabds}
\eea
where $r$ and $\bar r$ are the $F$-violating branching ratios defined
in section \ref{sec:cpviolation}.  To obtain the last line in
(\ref{eq:thetabds}), we have used (\ref{eq:chem1}) and
(\ref{eq:chem2}), $n_\chi = \int f_\chi d^3p/(2\pi)^3$, and the
thermal average of the $\chi$-decay width $\Gamma_\chi$,
\be
\langle \Gamma_\chi\rangle ={1 \over n_{\chi}} \int\!\!\Dx~ f_\chi \Gamma_\chi
= (\Gamma_\chi)_{\rm rest} {K_2(x) \over K_1(x)} ~.
\ee
 $K_1(x)$  and $K_2(x)$ are modified Bessel functions \cite{abram}, $x
 \equiv M/T$,  
and $(\Gamma_\chi)_{\rm rest}$ is the $\chi$-decay width  in the rest frame
$E_\chi = M$. We assume that $\Gamma_\chi$ 
is dominated by the renormalizable interactions, 
\be \label{eq:dominated}
\Gamma_\chi = {1\over 2E_\chi} \int \Do \Dt~
|M(\chi\rightarrow b_1d_2)|^2 \simeq
{g_1^2 M^2\over 8\pi E_\chi}~.
\ee

Another term in the Boltzmann equations corresponds to the
annihilation through 
renormalizable interactions $\chi\bar\chi \leftrightarrow b\bar b$ and its
back-process,
 \bea
\Theta (\chi\bar\chi \leftrightarrow b\bar b) 
& = & \int \Dx \DX \Db \DB
\left [ f_\chi f_{\bar\chi} |M(\chi\bar\chi \rightarrow b \bar b)|^2
\right. \nonumber\\
&& \left. \hspace{1.3in} - f_b f_{\bar b} |M(b\bar b \rightarrow \chi\bar\chi)|^2 
\right ] \nonumber\\
& = & \int \Dx \DX \Db \DB
\left [ f_\chi f_{\bar\chi} - f_\chi^{eq} f_{\bar\chi}^{eq}\right]
|M(\chi\bar\chi\rightarrow b\bar b)|^2~.
\nonumber\\
\label{eq:thetabb}
\eea
Here we have used the equality $|M(b\bar b \rightarrow \chi\bar
\chi)|^2 = |M(\chi\bar\chi \rightarrow b\bar b)|^2$, guaranteed by
$CPT$ invariance. This allows us to write (\ref{eq:thetabb}) in terms
of the $\chi \bar\chi$-annihilation cross section. In fact, using that
$f_b f_{\bar b} = f_\chi^{eq} f_{\bar\chi}^{eq}$, where
$f_{\bar\chi}^{eq}(p) = f_\chi^{eq}(p)$, and the definition of the
thermal average of $v \sigma$,
\be
\langle v\sigma (\chi\bar\chi \rightarrow b\bar b)\rangle =
{1 \over n_\chi n_{\bar\chi}}
\int \!\!\Dx \DX \Db \DB~ f_\chi f_{\bar\chi} |M(\chi\bar\chi
\rightarrow b \bar b)|^2~, 
\ee
 we get
\be
\Theta (\chi\bar\chi \leftrightarrow b\bar b) = \left(n_\chi n_{\bar\chi} -
(n_\chi^{eq})^2\right) \langle v\sigma (\chi\bar\chi \rightarrow b\bar
b)\rangle 
\ee
We can  always relate amplitudes for $3 \rightarrow 2$ reactions to
the corresponding amplitude for the 
$2 \rightarrow 3$  
back-process through CPT invariance. Therefore, proceeding as in this
example, only cross-sections for processes with two particles in the
initial state appear in the equations. In 
the same way $3 \rightarrow 1$ (or $2 \rightarrow 1$) processes are related to 
 $1 \rightarrow 3$ (or $1 \rightarrow 2$) decays.

Using the evolution equation for $n_{\bar\chi}$ analogous to
(\ref{eq:boltzmann}) 
(see appendix A),  
and changing to the dimensionless variable $x \equiv M/T$ through the
relation $dt = (xH)^{-1} dx$, we obtain,
\bea
{dY_+\over dx} = & - & {\langle\Gamma_\chi\rangle\over xH} 
\left [ (Y_+ - Y_+^{eq}) \right.\nonumber\\
&&\nonumber\\
& + & \left.\eta r_0 Y_F Y_+^{eq} - 2 r_0 Y_F Y_- Y_+^{eq} 
\right ]\nonumber\\
&&\nonumber\\
& - & 24 r_0  \frac{\langle \Gamma_\chi\rangle}{x^{3}H} 
\left [Y_+-Y_+^{eq}+Y_F Y_-\right ]\nonumber\\
&&\nonumber\\
& - & 96 r_0 \frac{\Gamma_{\chi {\rm rest}}}{x^{4}H} \left [ 2(Y_+-Y_+^{eq})
+ (Y_F -Y_F Y_+^{eq} + Y_-)Y_-\right ]\nonumber\\
&&\nonumber\\
& - & 2 \left [ Y_+^2 - (Y_+^{eq})^2 - Y_-^2 \right ] {n_\gamma\over xH}
\langle v\sigma (\chi\bar\chi\rightarrow b\bar b)\nonumber\\
& & \nonumber\\
& + & v\sigma'(\chi\bar\chi\rightarrow b\bar b\sigma) +
v\sigma' (\chi\bar\chi \rightarrow b\bar b \bar\sigma )\rangle +
\mbox{c.ch.}~~ .
\label{eq:y+}
\eea
As we will see in section 4, the first line of (\ref{eq:y+}) contains
the dominant terms 
and we will always be able to neglect the others.
The $1^{\mbox{\small st}}$ and $2^{\mbox{\small nd}}$ lines correspond
to $D$ and $ID$, the $3^{\mbox{\small rd}}$  
and $4^{\mbox{\small th}}$ to PS and the remaining lines to $RAN$, $NRAN$ 
and crossed channels (c.ch.) of the $NRAN$.
 
The prime in $\sigma'$ indicates that the contribution to the cross
section of a real intermediate particle ({\em i.e.,\/} an intermediate
particle on mass-shell) has been removed. Scattering or
annihilation processes involving a real intermediate particle are
already taken into account by other terms in the Boltzmann equations,
and must be subtracted to avoid double counting. 
For instance, production of a $\chi$ particle near the peak of the
resonance through inverse decay,
subsequently followed by its decay.  This kind of time-ordered
 sequence of processes is described by the terms
in the Boltzmann equation corresponding to each individual process.
In our example these are inverse decay and decay.

The need for subtraction of pole contributions is mentioned in earlier
papers on baryon-asymmetry generation \cite{kowo,frol}, but not
described in detail in the literature until recently \cite{clka},
after we had developed our own subtraction method. In order to
subtract the contribution of the pole, we compute the Laurent
expansion about $\Gamma_\chi=0$ of $\sigma$, or of the thermal average
$\langle\sigma v\rangle$ (since it may be easier to compute this
average with the complete cross section). We then subtract the term
proportional to $\Gamma_\chi^{-1}$, the only negative power occurring in
the expansion, and set $\Gamma_\chi = 0$.  We thus identify the term of
order $\Gamma_\chi^{0}$ as the virtual intermediate particle contribution
to the cross-section in the narrow-width approximation.  This
procedure is easily seen to be equivalent to the methods proposed in
\cite{clka}, to leading order in the coupling constants.  As pointed
out there, virtual cross-sections defined this way can take negative
values in the region around the pole.  This negative values have no
practical effects, as we have verified numerically, when the
subtracted cross-section appears in a damping term in the evolution
equations.

The case of $\chi\bar\chi$ annihilations, which are a source term and
in which the intermediate particle is a stable fermion, is discussed
in section \ref{sec:annbackpro}.

\subsection{Evolution Equation for $F$-Number Density}

The evolution of $n_\sigma$ is due to processes that involve
non-renormalizable interactions, the only ones that violate $F$-number
in our model.  These processes are non-renormalizable $\chi$ decays
$(NRD)$ and inverse decays $(NRID)$, ``point"-scatterings $(PS)$,
$\chi\bar\chi$ non-renormalizable annihilations ($NRAN$), their
crossed-channels $(NRCC)$ and 2 $\rightarrow$ 3 non-renormalizable
scatterings of $b, d $ and $\sigma $ $(NRS$ such as $bd\rightarrow
bd\sigma$, $b\bar d \rightarrow \bar bd\sigma$ etc.).

The evolution equation for $n_\sigma$ is of the form,
\be
{dn_\sigma\over dt} + 3 n_\sigma H = - \sum_\alpha \Theta_\alpha
\ee
where $\alpha = NRD$, $NRID$, $PS$, $NRAN$, $NRCC$, $NRS$.
The full equations for $n_\sigma$ and $n_{\bar\sigma}$ are given in
appendix A, and 
from them we obtain,
\bea
{dY_F\over dx} 
& = & 2r_0{\langle\Gamma_\chi\rangle\over xH} \left [ \eta (
Y_+ - Y_+^{eq}) - Y_F Y_+^{eq} -(1+2Y_+^{eq})Y_-\right ]\nonumber\\
& & -192 r_0 x^{-4} {{\Gamma_{\chi{\rm rest}}}\over H}
\left [ (Y_+ + Y_+^{eq}) Y_F +(2 + Y_+ + 3 Y_+^{eq}) Y_- \right ]
\nonumber\\
& & -48 r_0 x^{-3} {\langle \Gamma_\chi\rangle\over H}
\left [Y_F Y_+ +(1+ 2Y_+^{eq})Y_- \right ]\nonumber\\
& & + 4{n_\gamma \over xH} 
\left\{ \left [ Y_+^2 - (Y_+^{eq})^2-Y_-^2 \right ] 
\xi \langle v\sigma' (\chi \bar\chi \rightarrow b \bar b\sigma )\rangle 
\right. \nonumber\\
& & \left. + Y_F \left [ 2(Y_+^{eq})^2
\langle v\sigma'(\chi \bar\chi \rightarrow b \bar b \sigma)
+ v\sigma'(\chi \bar\chi \rightarrow b \bar b \bar\sigma)
\rangle \right.\right. \nonumber\\
& & \left.\left. + \langle v\sigma'(b d \rightarrow b d \sigma)
+ v\sigma'(\bar b \bar d \rightarrow \bar b \bar d \bar\sigma)\rangle 
+ {\rm c.ch.} \right ] \right\} \nonumber\\
& & + \mbox{more $2 \leftrightarrow 3$  c.ch.}\label{eq:yf}
\eea
The $1^{\mbox{\small st}}$ line corresponds to the non-renormalizable
decays and 
inverse decays, and some terms from $2 \leftrightarrow 3$ scatterings
needed to obtain the correct minus sign for $Y_+^{eq}$, as explained below.
The $2^{\mbox{\small nd}}$ and $3^{\mbox{\small rd}}$ lines
correspond to PS processes. 
The rest of (\ref{eq:yf}) corresponds to non-renormalizable annihilations and 
the remaining terms of $2 \leftrightarrow 3$ scatterings (c.ch. stands for
crossed channels).

If we take into account only decays and inverse decays and neglect
other processes, the r.h.s.\ of (\ref{eq:yf}) reduces to,
\be \label{eq:wrongsource}
2\eta r_0 \frac{\langle\Gamma_\chi\rangle}{xH} (Y_+ + Y_+^{eq})~~.
\ee 
Notice the
plus sign in front of $Y_+^{eq}$. It is easy to see that a minus sign
is  necessary, because no $F$-number should be generated if $Y_+
= Y_+^{eq}$, {\em i.e.,\/} if there is no departure from equilibrium.
 The correct sign is obtained once the  $2 \leftrightarrow 3$
scattering  terms proportional to
$|M'(bd\rightarrow bd\sigma)|^2 -  |M'(\bar b\bar d \rightarrow \bar b \bar
d \bar \sigma)|^2$ are included.  The unitarity relation,
\[
\sum_i|M(i\rightarrow j)|^2 = \sum_i |M(\bar{\imath} \rightarrow j)|^2
\]
applied to $|j\rangle= |bd\sigma \rangle$, relates scatterings and decays,
\bea
&& \int \left [\Dx |M(\chi \rightarrow bd\sigma)|^2
+ \Do \Dt |M'(b d \rightarrow bd\sigma)|^2 \right ]  
\nonumber\\
& =& \int \left [\DX |M(\bar\chi \rightarrow \bar b \bar d\bar\sigma)|^2
+ \Do \Dt |M'(\bar b \bar d \rightarrow \bar b\bar  
 d\bar\sigma)|^2 \right ]~~.  
\label{eq:decscatt}
\eea
Thus, the above-mentioned scattering terms are proportional to the 
non-renormalizable decay rates, namely
\bea
&& \int \Do \Dt \left [ |M'(b_1d_2\rightarrow b d \sigma)|^2
-M'(\bar b_1 \bar d_2\rightarrow \bar b \bar d \bar\sigma)|^2 \right ]
\nonumber\\
& =& \int \Dk \left [ 
|M(\bar\chi_*\rightarrow \bar b\bar d\bar\sigma)|^2
-M(\chi_*\rightarrow bd\sigma)|^2 \right ]~~.
\eea
Therefore, the contribution of $2\leftrightarrow 3$ scatterings to the
source terms on the r.h.s.\ of (\ref{eq:yf}) is proportional to
$Y_+^{eq}$ and is given by,
\bea
& &- {2\over xHn_\gamma}(1+2Y_-Y_F)
\int\!\!\Do \Dt \DT \Df \DF~ e^{-(E_1+E_2)/T} 
\nonumber\\
& &\hspace{.4in} \left[ |M'(b_1 d_2 \rightarrow b_3 d_4 \sigma_5)|^2 
-|M'(\bar b_1 \bar d_2 \rightarrow \bar b_3 \bar d_4 \bar\sigma_5)|^2 \right] \nonumber\\
& = &- {2\over xHn_\gamma}(1+2Y_-Y_F)
\int\!\!\Do \Dt \DT \Df \DF~ e^{-(E_3+E_4+E_5)/T}
\nonumber\\
& &\hspace{.4in} \left [ |M'(b_1d_2\rightarrow b_3d_4\sigma_5)|^2
-|M'(\bar b_1\bar d_2\rightarrow \bar b_3 \bar d_4\bar\sigma_5)|^2 \right]
\nonumber\\
& = & - {2\over xHn_\gamma}(1+2Y_-Y_F)
\int\!\!\Dk \DT \Df \DF~ e^{- E_* /T}
\nonumber\\
& &\hspace{.4in} \left[ |M(\bar\chi_* \rightarrow \bar b_3 \bar d_4 \bar\sigma_5)|^2 -|M(\chi_* \rightarrow b_3 d_4 \sigma_5)|^2 \right ] 
\nonumber\\
& = & - {\langle \Gamma_\chi \rangle \over xH}
4\eta r_0 Y_+^{eq}(1+2Y_-Y_F) ~~,
\label{eq:scatt}
\eea
which together with (\ref{eq:wrongsource}) adds up to the source term
on the first line of (\ref{eq:yf}).

Notice that in (\ref{eq:yf}), besides $\eta$ that parametrizes the
violation of $CP$ in decays , there
appears also the parameter $\xi$ for the $CP$-violation in
annihilations 
(see equations (\ref{eq:eta}) and (\ref{eq:xi})).

\subsection{Parameters and Initial Conditions}
\label{sec:paramandincond}

In order to solve numerically the coupled evolution equations for $Y_+$ 
and $Y_F$ we impose two initial conditions at a small enough value of 
$x = x_0$, such that at $x_0$ it can be safely assumed that $\chi$
and $\bar \chi$  are in equilibrium.  Thus, we take
\be
Y_+(x_0) = Y_+^{eq} (x_0), ~~~Y_F(x_0) = Y_{F0}~,
\ee
where $Y_{F0}$ is equal to zero only  for $F$-symmetric initial
conditions. Numerically, any $Y_{F0} \ll Y_F(\infty) \simeq 10^{-10}$
will be equivalent to zero,
and in most cases we take $Y_{F0} = 10^{-20}$ as our $F$-symmetric initial
condition. We have also studied $F$-asymmetric initial conditions
(section \ref{sec:asymm}), assuming that some other F-generating processes have
acted at  earlier times.
In this case, for large enough $Kr_0$ there is an early erasure and 
subsequent generation of $Y_F$ due to the processes we consider here.

Because our effective Lagrangian is not valid at energy scales
larger  than $\Lambda$, we restrict our choice of $x_0$ to values where $T
<  \Lambda$.  At any rate, the solutions of the evolution
equations are stable against variations in $x_0$, as long as $x_0$ is not too 
close to 1, $x_0 \lesssim 0.1$, so
\be\label{eq:x0compelling}
0.1 \gsim x_0 \gsim {M\over \Lambda} \simeq 30 {g_1\over g_2}
\sqrt{r_0}~~. 
\ee
In the absence of any compelling reason to do otherwise, we
take  $g_1 / g_2\sim 1$. 

The evolution of the $F$-asymmetry in our model is determined by five
independent parameters (besides the initial conditions $x_0, Y_{F0})$.
It is clear from the evolution equations (\ref{eq:y+}) and
(\ref{eq:yf}) that the most suitable parameters to classify different
scenarios for the production of $Y_F$ are: $K$, the ``effectiveness of
decay'' parameter, $Kr_0$, the ``effectiveness of 
non-renormalizable reactions" parameter, $g_1$ the 
coupling constant of renormalizable interactions, and $\eta$ and
$\xi$, the only two independent 
CP-violation parameters.  Therefore, we use
\be
 K, r_0, g_1, \eta~~ {\rm and}~~ \xi~,
\ee
as independent parameters in the evolution equations. For our numerical
solutions we choose reasonable but arbitrary values for three of them,
\be \label{eq:arbitrary}
g_1 = 10^{-1}~,~~~~\xi = \eta \simeq 5~10^{-4}~,
\ee
and examine different ranges of values of $K$ and $Kr_0$.
 
$K$ is the parameter whose value determines different baryon-asymmetry
scenarios in out of equilibrium decay models with renormalizable
interactions,
\be \label{eq:kexact}
K \equiv {(\Gamma_\chi)_{\rm rest}\over H(x=1)} = {1\over{\sqrt{g_*}}}
\biggl [ {g_1^2\over 8\pi} + {g_2^2\over 192(2\pi)^3}
\biggl( {M\over \Lambda}\biggr )^2 \biggr ] {M_P\over M}.
\ee
Here $(\Gamma_\chi)_{\rm rest}$ is the decay rate of non relativistic $\chi$
particles, $\Gamma_\chi (T \gsim M) = (\Gamma_\chi)_{\rm rest}$ (it is
the decay rate in the rest frame where $E_\chi =M$), $g_*$ is the 
effective number of relativistic degrees of freedom entering into the 
Hubble constant, that we take to be $g_* = 100$, and $M_P$ is the Planck 
mass.  We assume that the decay width of $\chi$, $\Gamma_\chi$, 
is dominated by the renormalizable decays (see 
(\ref{eq:dominated})), namely the first 
term in (\ref{eq:kexact}) is dominant,
\be \label{eq:kapprox}
K \simeq {g_1^2\over \sqrt{g_* 8\pi} } {M_P\over M} .
\ee
With this assumption on $\Gamma_\chi$, the parameter $r_0 \simeq 
\Gamma_\chi^{NR}/\Gamma_\chi$  defined in  (\ref{eq:r0}), is
\be \label{eq:r0approx}
r_0 \simeq {1\over 192\pi^2} \biggl ( {g_2 M\over 
g_1\Lambda}\biggr )^2~.
\ee
 $K$ and $Kr_0$ determine the 
``effectiveness of the reactions'' because all terms in $(dY_+/dx)$ in 
(\ref{eq:y+}) are proportional to $K$ (even the last one, where $n_\gamma
\langle v\sigma\rangle/H \sim K g_1^2/x$) and all terms in
$(dY_F/dx)$ in (\ref{eq:yf})  are proportional to $Kr_0$ (even the last one, 
where $n_\gamma \langle v\sigma\rangle/H \sim Kr_0 g_1^2/x$).  As we 
will see, the final value of $Y_F$, $Y_F(x\rightarrow\infty)$, depends 
only on $r_0$ for $K \lesssim 1$ and on $K$ and $r_0$ for $ K\gsim 
10$ 

\section{ Evolution of $Y_+$}
\setcounter{equation}{0}

The evolution of $Y_{+}$ is mainly determined by the first term in
(\ref{eq:y+}), due to decays and inverse decays.  We can then write,

\be \label{eq:y+approx}
{dY_+\over dx} \simeq -~ Kx {K_1(x) \over K_2(x)} (Y_+ -
Y_+^{eq}). \ee
%

In first approximation, we shall set $Y_{F}=0$, $Y_{-}=0$ in
(\ref{eq:y+}).  This approximation will be justified below. 

\begin{figure}[t]
  \begin{center}
    \leavevmode
    \begin{picture}(300,175)(100,50)
    \put(100,0){\epsfbox{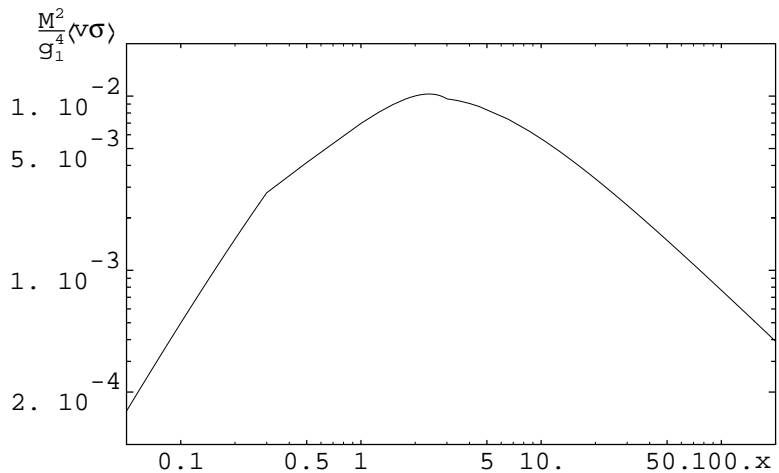}}
  \end{picture}
  \vspace{-4 ex}
  \parbox{300pt}{
    \protect\caption{Cross-section for renormalizable annihilations
      $\chi\bar{\chi}\rightarrow b\bar{b}$ as a
      function of x.}
    \protect\label{fig:fit2renann}}
  \end{center}
\end{figure}

\begin{figure}[t]
  \begin{center}
    \leavevmode
    \begin{picture}(300,200)(100,50)
    \put(100,0){\epsfbox{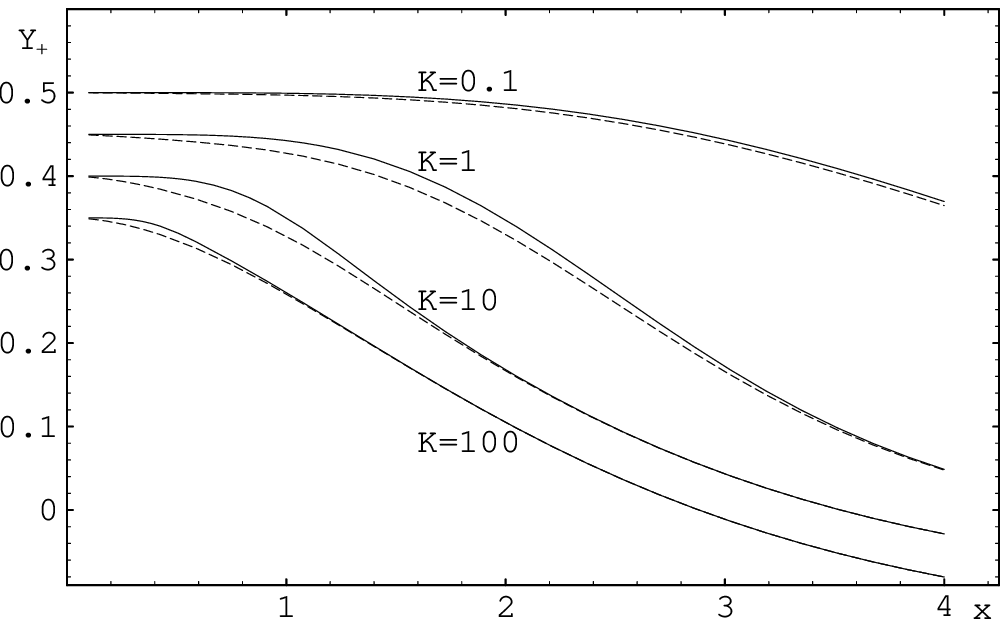}}
    \end{picture}
    \parbox{300pt}{
    \protect\caption{$Y_+ (x)$ for various values of $K$. 
      The curves have been vertically diplaced
      for clarity.  Dashed lines correspond to solutions to the full
      evolution equation, and solid lines to the simplified equation
      (see text).}
    \protect\label{fig:yplusplot}}
  \end{center}
\end{figure}

The $5^{\rm th}$ and $6^{\rm th}$ lines in (\ref{eq:y+}), correspond
to $\chi\bar\chi$ annihilations.
Non-renormalizable annihilation terms are always much smaller than the
renormalizable one, on the $5^{\rm th}$ line of (\ref{eq:y+}), and can be
neglected.  This is due to the fact that $(g_2 M/g_1\Lambda)^2 \ll 1$,
and also to  phase-space considerations.  The cross-section for
$\chi\bar\chi\rightarrow b\bar{b}$ can be expressed as,
\be
\langle v\sigma(\chi
\bar \chi \rightarrow b \bar b)\rangle = {g_1^4 \over M^2} f(x)~~. 
\ee
A fit to the function $f(x)$ is shown in figure \ref{fig:fit2renann}.
At small $x\ll 1$ this term is dominant.  However, in this region
$Y_{+}$ closely follows $Y_{+}^{eq}= 1/2 - O(x^2)$, the difference
being of order $Y_{+}-Y_{+}^{eq}\sim O(x^2)$.  Thus, $Y_{+}$ remains
approximately constant, independently of the detailed form of the
r.h.s.\ of (\ref{eq:y+}).  A plot of $Y_+(x)$ for several values of
$K$ in figure \ref{fig:yplusplot} shows the effect of the annihilations
term which, as argumented, is small. 

For $x\gsim 1$, the annihilation rate is suppressed relative to
decays, and the first term in (\ref{eq:y+}) determines the evolution
of $Y_{+}$.  It is in this region, in which $\chi$ particles are becoming
non-relativistic and $Y^{eq}_{+}$ varies rapidly, where large
departures from equilibrium can occur.  This is, then, the most
relevant region for generating an asymmetry $Y_{F}$.

The ratio of the $5^{\rm th}$ line, to the $1^{\rm st}$ term, contains 
$(Y_+ +Y_+^{eq} )\lesssim 1$,  and the
factor $n_\gamma\langle v\sigma\rangle  /\langle \Gamma_\chi\rangle = 32
g_1^2 K_2(x) f(x)/K_1(x) x^3$, where $f(x)$, the fit function just
mentioned, is  $f(x) < 10^{-2}$. 
Thus, for values of $x$ not much smaller than 1, the $5^{\rm th}$ line of
(\ref{eq:y+}) is smaller than the $1^{\rm st}$  term by a factor $g_1^2
K_2(x)/x^3 K_1(x)$ $\ll 1$. 

We also have that the ratio of the $1^{\rm st}$ term to the one on the
 $3^{\rm rd}$ line of (\ref{eq:y+}) is given by $x^2/24 r_0\gg 1$, 
 for $x\gsim 1$.  A similar argument holds for the term
 on the $4^{\rm th}$ line.  Both terms come from ``point''-scattering
 processes. 

 We have numerically checked that $Y_-$ is always at least one order
of magnitude smaller than $Y_F$ and $Y_+$, so it can be safely
ignored. This can be explained by noticing that  the source term in the
evolution equation for $Y_-$ (see appendix A) is proportional to $r_0 Y_F
Y_+$ and  $r_0 \ll 1$.

We are then left with the term proportional to $\eta r_0 Y_F$ on the
$2^{\rm nd}$ line in (\ref{eq:y+}), stemming from non-renormalizable
decays and inverse decays.  This term would be non-negligible only for
very large values of $Y_F$
\be
Y_F \simeq {1 \over \eta r_0} {{(Y_+ - Y_+^{eq})}\over
Y_+^{eq}},
\ee
that would never be produced through the mechanism discussed in this
paper.  These large values of $Y_F$ could only arise 
from large initial values $Y_{F0}$ of the $F$-asymmetry, if its erasure, as
discussed  below, is not efficient.  However in this case the value of
$Y_F$ never  significantly departs from $Y_{F0}$ and the mechanism discussed
here becomes  irrelevant.

We mentioned above that we only need to find the density of $\chi$
particles, $Y_+$, produced 
at $x$ of order 1 and larger in order to estimate $Y_F$.
Let us consider this statement in more detail.  We shall treat
separately the cases of $K \ll 1$ and $K \gsim 1$. 

For $K \ll 1$, all $\chi$-number changing interactions are out of 
equilibrium at $x \simeq 1$.
Thus $Y_+$ does not change with respect to $Y_+(x\simeq 1) \simeq Y_+^{\ 
eq}(x\simeq 1) \simeq 0.5$ until the $\chi$-decays occur, at 
$x\simeq x_{\rm Decay}$.   
We define  $x_{\rm Decay}$ as the value of $x$ at which $\chi$-decays
enter thermal  equilibrium, namely,
\[ 
\langle \Gamma_\chi(x_{\rm Decay})
\rangle = H (x_{\rm Decay}) = H(x=1)/x_{\rm Decay}^2~~,
\]
and using $\langle \Gamma_\chi(x_{\rm Decay}) \rangle \simeq
(\Gamma_\chi)_{\rm rest}$ we get $x_{\rm Decay} \simeq (\sqrt K)^{-1}$. 
Since $dY_F/dx \sim Y_+ - Y_+^{eq}$, the overabundance of $\chi$ in this
case is responsible for  most of the $Y_F$ produced. 

 For $K \gsim 1$, instead,
there is never a  large overabundance of $\chi$, because the reactions
that change  $\chi$-number are in equilibrium at $x \simeq 1$.
While $\chi$-decays are always in equilibrium, inverse decays and 
$\chi$-number generating scatterings go
out of equilibrium at  some point $x_f$ at which all
remaining $\chi$-particles decay. Therefore, because
 $dY_F/dx \sim Y_+-Y_+^{eq}$, the production of
$Y_F$  happens steadily for both $x \lesssim 1$
and $x \gsim 1$ and, consequently, the  production for $x$ of $ O(1)$ and
larger gives the right order of  magnitude\footnote{
To clarify this argument
assume the production rate is almost constant, i.e. $dY_F/dx \simeq C$
where $C$ is a  constant, so the increment of $Y_F$ is proportional to
the increment of $x$,  $\Delta Y_F
\simeq C \Delta x$ and $Y_F(\infty) \simeq C  x_f$, since
$\Delta x \simeq  x_f$ for $ x_f \gg1$  while the
production of $Y_F$ that happens before $x \simeq 1$ can  contribute at
most $\Delta Y_F \simeq C$, which does not change the order  of magnitude
of $Y_F(\infty)$.}.

An approximate analytical solution of (\ref{eq:y+approx}), for $K \geq
1$ and all  
$x$ is given in appendix C.  We will actually only use a simpler 
approximation to this solution, valid for $K \gsim 10$ and $K^{-1/3}
\ll x $,
\be \label{eq:deltapprox}
{Y_+ (x) - Y_+^{eq}(x) \over Y_+^{eq}} \simeq {1\over Kx}
\ee

\section{ Evolution of $Y_F$: Symmetric Initial Conditions.}
\label{sec:yfevolsym}
\setcounter{equation}{0}

In this section we give analytical and numerical estimates of the
asymmetry $Y_F$ 
produced in different regimes, characterized by small or large values
of $K$ and $Kr_0$.  We shall be more specific about these regimes
below.

The evolution equation (\ref{eq:yf}) has two source terms,
corresponding to generation of $Y_F$ by decays and annihilations of
$\chi$ particles, respectively, and their inverse processes.  We shall
consider them separately, deferring the treatment of annihilations
until the end of this section since, as we shall see,  they turn out
to give a negligible contribution compared to decays. 

\subsection{Decays and Inverse Decays as a Source.}
\label{sec:decays}

For $K\lesssim 1$, all 
 $\chi$-number changing processes are out of equilibrium at $x \gsim 1$.
$Y_+$ remains constant beyond $x=1$,
\[
Y_+(x > 1)
\simeq Y_+(x=1)\simeq Y_+^{eq}(x=1)= 0.5~~, 
\]
until $\chi$ and
$\bar \chi$ decay.  This happens when the  decay rate finally equals
the expansion rate 
of the Universe, $\langle\Gamma_\chi(x_{\rm Decay})\rangle = H(x_{\rm
  Decay})$, at 
$x_{\rm Decay} \simeq K^{-1/2}$. This overabundance of $Y_+$ for $x\gsim
1$ is responsible for most of the  $Y_F$ produced.  In this case it is easy
to estimate the final $F$-asymmetry, 
because each $\chi \bar\chi$ pair that decays produces
 a net $F$-number 
\[
F_{\bar\sigma} (\bar r-r) = 
F_{\bar\sigma} 2\eta r_0~~,
\]
where the CP-violating parameter $\eta$ was defined in (\ref{eq:eta})
and (\ref{eq:r0}).   
Thus, we obtain
\be \label{eq:yfinf}
Y_F(\infty)\simeq F_{\bar\sigma} 2\eta
r_0 Y_+(x_D)= \eta r_0~,
\ee
 since $Y_+(x_D)\simeq 1/2$, and we have effectively set  
$F_{\bar\sigma}=1$ for simplicity (see (\ref{eq:yfs})).

For $K > 10$, $\chi$-number changing reactions are in equilibrium at
$x \gsim 1$, therefore $Y_+(x > 1)$ follows $Y_+^{eq}$ closely.  We
have to consider different regimes characterized by the value of
$Kr_0$.  As shown below, for $\xi\simeq\eta$ and $g_1^2 \leq 10^{-2}$
(corresponding to our choice of parameters in (\ref{eq:arbitrary})),
the annihilation term is negligible as a source for $Y_F$ compared to
the decay term for reasonable values of $K$ and $Kr_0$.  Thus, taking
into account only decays, we have the following results.
\begin{itemize}
\item[*] For $Kr_0 < 10^{-2}$, we have $Y_F(\infty) \sim  r_0\eta$.  Notice
that $r_0 \lesssim 10^{-3}$ (see (\ref{eq:r0approx})), so this case includes
in particular the case $K<1$.
\item[*] For $10^{-2}<Kr_0\lesssim 10$ we have damping by point scatterings,
\[ Y_F(\infty) \sim {\eta[\ell n(10^2 Kr_0)]^2\over 10^2 K}~~.\]
\item[*] For $10\lesssim Kr_0\lesssim 10^7$ inverse decays are the dominant
damping process and we have, 
\[Y_F(\infty) \sim {\eta\over [\ell n(10^2  Kr_0)] K}~~.\] 
\item[*] For very large values of $Kr_0$, $Kr_0\gsim 10^7, $ we obtain
exponential damping of the form $Y_F(\infty) \sim \exp\{-(36\pi
  Kr_0g_1^2)^{1/6}\}$, due to $2\leftrightarrow 3$ scatterings.
\end{itemize}
In the following subsections we discuss these results in more detail.

\subsection{Damping by point scatterings}
\label{sec:dampingbypointscatterings}

We have seen that for $Kr_0 \lesssim 10^{-2}$ we can ignore damping
terms.  As $Kr_0$ grows larger other terms in the equation for $Y_F$
start being relevant.  We shall take them into account in what follows,
beginning with the term corresponding to point scattering processes.
In all cases we evaluate $Y_F$ by quadratures, applying the
saddle-point approximation where appropiate.  

We shall now assume $192 K r_0\gsim 1$.  The evolution equation
including only the source term due to decays and the damping term due
to point scatterings reads,
\bea\label{eq:pointscat1}
\frac{dY_F}{dx} & = & 2\eta r_0 K \frac{K_1(x)}{K_2(x)} x
\Delta(x)\nonumber\\
  &  & \mbox{}-192 K r_0 \frac{1}{x^2}\left(\rule{0ex}{2ex}Y_+ +
    Y_+^{eq}\right) Y_F~~ 
\eea
with,
\[
\Delta (x) \equiv Y_+ (x) - Y_+^{eq} (x).
\]

The expression for $Y_F(\infty)$ can then be written as,
\bea
Y_F (\infty) & = & \int_{x_0}^\infty \!\! du ~ 2\eta K r_0
\frac{K_1(u)}{K_2(u)} u \left[ \Delta(u) e^{\displaystyle u} \right]
\times\nonumber\\
&  & \exp \left\{ -u-\int_u^\infty \!\! dz ~ \frac{192
    K r_0}{z^2} \left(\rule{0ex}{2ex}Y_+(z) + Y_+^{eq}(z)\right)
\right\}~~. 
\label{eq:formalsol}
\eea
Since we are considering decays as the source for $Y_F$, we expect
this integral to be dominated by the contribution of the region $u>1$.
For this reason we explicitly extracted the exponential dependence
from $\Delta$.  Furthermore, we have $K\gsim 1/(192 r_0)$, so we can
use the leading-order approximate expressions,
\bea\label{eq:leadingorder}
\Delta (u)  \simeq  \frac{Y_+^{eq}(u)}{K u} & ; & 
Y_+(u) + Y_+^{eq}(u)  \simeq  2 Y_+^{eq}(u) \\
\frac{K_1(u)}{K_2(u)}  \simeq  1 & ; &
Y_+^{eq} (u) \simeq  \sqrt{\frac{\pi}{2}} \frac{u^\frac{3}{2}}{4}
e^{\displaystyle -u}~,
\eea
to evaluate $Y_F(\infty)$.

To leading order in $K$ and $u$ we then have,
\be\label{eq:leading}
Y_F(\infty) = \sqrt{\frac{\pi}{2}} \frac{\eta r_0}{2} \int_{x_0}^\infty
u^{\frac{3}{2}} e^{\displaystyle  {\cal E} (u)}~,
\ee
where we defined the exponent ${\cal E}(u)$ as,
\be\label{eq:exponent}
{\cal E}(u) =  u + 96 K r_0 \sqrt{\frac{\pi}{2}} \Gamma
\left(\frac{1}{2},u\right)~.
\ee
Writing $\cal E$ in terms of an incomplete Gamma
function \cite{abram} will be useful below, when
we consider more terms in the equation.

The exponent (\ref{eq:exponent}) is minimal at $u=u_F$ given by the
equation
\be\label{eq:uf3}
\sqrt{u_F} e^{\displaystyle u_F} = 96 \sqrt{\frac{\pi}{2}} K r_0~,
\ee
corresponding to the epoch of ``freeze-out'' of the damping process
\cite{kotu}.  At the minimum we have,
\bea
{\cal E}(u_F) 
& \simeq & u_F + 1\\
{\cal E}^{\prime\prime}(u_F) & = & 1 + \frac{1}{2u_F}~.
\eea
We then obtain the expression,
\be\label{eq:adfa}
Y_F(\infty) = \frac{\sqrt{2\pi}}{192 e} \frac{\eta}{K}
\frac{u_F^2}{\sqrt{1+\frac{\displaystyle 1}{\displaystyle 2 u_F}}}~,
\ee
for the final asymmetry.  

The freeze-out epoch $u_F$ is an increasing function of $K r_0$.  We
can then roughly approximate, 
\be
u_F \simeq \ell n \left(96 \sqrt{\frac{\pi}{2}} K r_0\right)
\ee
for large enough values of $K r_0$.  In this way we arrive at an
explicit form for $Y_F$, 
\be\label{eq:pointscatfinal}
Y_F(\infty) = \frac{\sqrt{2\pi}}{192 e} \frac{\eta}{K}
\left[ \ell n \left(96 \sqrt{\frac{\pi}{2}} K r_0\right)\right]^2~,
\ee
where we ignored the square root in the denominator of
(\ref{eq:adfa}).

Notice that $Y_F(\infty)$ in (\ref{eq:pointscatfinal}) depends on $r_0$
only through the combination $Kr_0$, and that the dependence is very
weak.  This is unlike the ``free decay'' case where we had a linear
dependence on $r_0$.   The transition between the two regimes is, of
course, not sharp.  The linear dependence with $r_0$ becomes flatter
as $Kr_0$ grows, turning into the logarithmic form of
(\ref{eq:pointscatfinal}) at about $Kr_0\simeq 10^{-2}$.

The factor of $K$ in the denominator
represents the damping due to point scatterings.  This suppression of
the generated asymmetry as a power of $K$ is similar to the damping
obtained in renormalizable models \cite{kotu}.

\subsection{Damping by inverse decays}
\label{sec:inversedecays}

The next term we shall take into account corresponds to inverse
decays.  It is the term proportional to $Y_F$ on the first line of
(\ref{eq:yf}).   The evolution equation now reads,
\bea\label{eq:terms3+2}
\frac{dY_F}{dx} & = & 2r_0 K \frac{K_1(x)}{K_2(x)} x
\left(\rule{0ex}{2ex}\eta\Delta(x)-Y_F Y_+^{eq}\right)\nonumber\\
  &  & \mbox{}-192 K r_0 \frac{1}{x^2}\left(\rule{0ex}{2ex}Y_+ +
    Y_+^{eq}\right) Y_F 
\eea

We shall proceed as in the previous section.  $Y_F(\infty)$ is given
by (\ref{eq:leading}), but now there is an extra term in the exponent,
\be
{\cal E}(u) = u + \frac{K r_0}{2} \sqrt{\frac{\pi}{2}} \left( 192\Gamma
\left(\frac{1}{2},u\right) + \Gamma \left( \frac{7}{2},u\right)\right)~.
\ee
The equation for the freeze-out point $u_F$ takes the form,
\be\label{eq:uf2+3}
\sqrt{\frac{\pi}{2}}\frac{Kr_0}{2} \left( \frac{192}{\sqrt{u_F}} +
  u_F^{\frac{5}{2}}\right) e^{\displaystyle -u_F} =1~.
\ee
When $u_F^{5/2} < 192/\sqrt{u_F}$, corresponding to $K r_0 \lesssim 6.5$, we
can neglect the second term in the parentheses in (\ref{eq:uf2+3}).
In this case we recover the results from the previous section.

For $K r_0 \gsim 6.5$ we consider the approximate equation for $u_F$
(compare with (\ref{eq:uf3})),
\be\label{eq:appsaddle}
\sqrt{u_F} e^{\displaystyle u_F} =
u_F^3\sqrt{\frac{\pi}{2}}\frac{Kr_0}{2} ~~.
\ee
At $u=u_F$ the exponent is minimal, taking the value,
\bea
{\cal E} (u_F) 
& \simeq & u_F + 1 \\
{\cal E}^{\prime\prime} (u) 
& \simeq & 1 - \frac{5}{2} u_F^{-1}~,
\eea
where 
we used $5 u_F^3 > 192$, and (\ref{eq:appsaddle}).

Using the approximate saddle-point condition (\ref{eq:appsaddle}), we
then obtain the expression for $Y_F(\infty)$,
\be\label{eq:adgb}
Y_F(\infty) = \frac{\sqrt{2\pi}}{e}\frac{\eta}{Ku_F}
\frac{1}{\sqrt{1-\frac{\displaystyle 5}{\displaystyle 2u_F}}}
 ~~~\mbox{for}~~~ Kr_0 > 6.5~~,
\ee
to be compared with (\ref{eq:adfa}).

In order to obtain an explicit expression for $Y_F$ we need a solution
to (\ref{eq:appsaddle}).  We cannot neglect the factor $u_F^3$ in that
equation, because that would be inconsistent with our previous
approximations.  Instead, for moderate values of $K r_0 \gsim 6.5$ we
can use the matching condition with the regime of the previous section
and write
\[
u_F \simeq \ell n\left( 96\sqrt{\frac{\pi}{2}} K r_0\right)~~,
\]
to obtain,
\be
Y_F(\infty) = \frac{\sqrt{2\pi}}{e}\frac{\eta}{K~\ell n\!\left(
    96\sqrt{\frac{\pi}{2}} K r_0\right)} ~~.
\ee
We see that the damping effects of inverse decays are stronger than
those of point scatterings, and that they appear at a later epoch.

For larger values of $Kr_0$, $K r_0\gg 6.5$, we must use an iterated
solution of (\ref{eq:appsaddle}),
\[
u_F \simeq \ell n \! \left( \frac{1}{2} \sqrt{\frac{\pi}{2}} K r_0 \right)
+ 3 \ell n \! \left( \ell n \! \left( \frac{1}{2} \sqrt{\frac{\pi}{2}}
    K r_0 \right)\right) 
\]
and replace it in (\ref{eq:adgb}).

\subsection{Point scatterings revisited}

Point scattering processes give rise to another damping term in the
evolution equation, given by the $3^{\rm rd}$ line of (\ref{eq:yf}).
Repeating the same analysis as in the previous sections, we are led to
the following expression for the exponent,
\be
{\cal E}(u) = u + \frac{K r_0}{2} \sqrt{\frac{\pi}{2}} \left( 192~\Gamma\!
\left(\frac{1}{2},u\right) + \Gamma\! \left( \frac{7}{2},u\right)+
24~\Gamma\! \left( \frac{3}{2},u\right)\right)~. 
\ee
The last term in $\cal E$ is new.  It never dominates the exponent,
however, and only gives small corrections to $Y_F(\infty)$, not larger
than 30\%.  We will, therefore, not take it into account, since we are
interested only in order-of-magnitude estimates.

\subsection{Damping by $2\leftrightarrow3$ scatterings}

Having analyzed the effect of the damping terms corresponding to point
scatterings and inverse decays, we are left with those related to
$2\leftrightarrow 3$ scatterings.  The number of diagrams for this
kind of processes is large, making the analysis quite intrincate.

Since the simplicity of our model does not warrant a detailed
treatment of this problem, we shall consider only two examples which
we consider representative of the general situation.  As we shall see,
this processes turn out to be relevant only for very large values of
$Kr_0 \gsim 10^7$, which for fixed $r_0$ correspond to $K\gsim
10^{10}$.

\begin{figure}[htbp]
  \begin{center}
    \leavevmode
    \begin{picture}(100,75)(100,90)
      \put(-50,0){\epsfbox{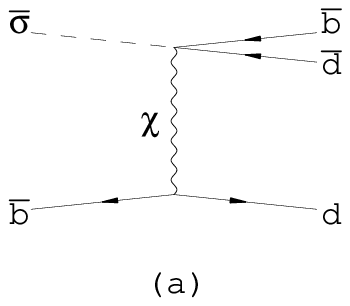}}
      \put(90,0){\epsfbox{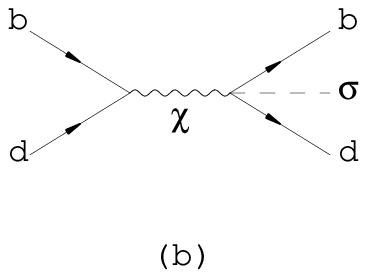}}
    \end{picture}
    \caption{Two diagrams for $2\rightarrow 3$ scattering.}
    \label{fig:two2three}
  \end{center}
\end{figure}

The first reaction channel we shall consider is
$\bar{b}\bar{\sigma}\rightarrow\bar{b}\bar{d} d$ (see figure
\ref{fig:two2three}a).  Its thermally averaged cross-section, 
for small and large values of $x$, has the asymptotic form,
\renewcommand{\arraystretch}{1.5}
\begin{equation}
  \label{eq:tasym}
  \langle v \sigma\rangle \simeq \left\{
    \begin{array}{ll}
      \displaystyle G^2~ \frac{0.012}{x^4} & x\gsim 10\\
      \displaystyle G^2 (2.83\times 10^{-5} - 1.27\times 10^{-4}\ell
      n(x^2)) & x\lesssim 0.1
    \end{array}
    \right.~~,
\end{equation}
\renewcommand{\arraystretch}{1}  
where we used the notation $G^2\equiv g_1^2 g_2^2/\Lambda^2$ for
brevity.

Adding the corresponding term to (\ref{eq:terms3+2}) we obtain,
\bea\label{eq:terms3+2+8}
\frac{dY_F}{dx} & = & 2r_0 K \frac{K_1(x)}{K_2(x)} x
\left(\rule{0ex}{2ex}\eta\Delta(x)-Y_F Y_+^{eq}\right)\nonumber\\
  &  & \mbox{}-192 K r_0 \frac{1}{x^2}\left(\rule{0ex}{2ex}Y_+ +
    Y_+^{eq}\right) Y_F \nonumber\\
  &  & \mbox{}-2^{11} 3 \pi Kr_0 \frac{g_1^2}{x^2}~\frac{\langle
    v\sigma\rangle}{G^2}~Y_F(x)~~.
\eea
For small values of $x$, taking into account (\ref{eq:tasym}), the
last term can be written as,
\begin{equation}
  \label{this}
  2^{11} 3 \pi Kr_0 \frac{g_1^2}{x^2}~\left(2.83\times 10^{-5} - 1.27\times
  10^{-4}\ell n(x^2)\rule{0ex}{2ex}\right) ~Y_F(x)~~,
\end{equation}
and turns out to be negligible compared to the term on the second line
of (\ref{eq:terms3+2+8}) for $x> 10^{-10}$.  The latter will,
therefore, be the most relevant term in determining $Y_F(\infty)$ for
$Kr_0 \lesssim 6.5$, as we have seen in previous sections since, for 
 reasonable values of the parameters, $x_0\gg 10^{-10}$.

For very large values of $Kr_0\gg 6.5$, $2\leftrightarrow 3$
scatterings can be important.  In this regime, we write $Y_F(\infty)$
in the form (\ref{eq:leading}), with the exponent given by
\begin{eqnarray}
  \label{2to3exp}
  {\cal E} (x) & = & x + Kr_0 \int_{x}^{\infty}\!\! dz ~ \left[
    2 z Y_{+}^{eq}(z) + 2^{11} 3 \pi \frac{g_1^2}{z^2}~\frac{\langle
      v\sigma\rangle}{G^2}\right] \\
  {\cal E}^\prime (x)& = & 1 - Kr_0 \left[
    2 x Y_{+}^{eq}(x) + 2^{11} 3 \pi \frac{g_1^2}{x^2}~\frac{\langle
      v\sigma\rangle}{G^2}\right] 
\end{eqnarray}
The minimum of $\cal E$ is found numerically to satisfy $x_F > 10$, so
we can approximate the expression for $\cal E^\prime$ as,
\begin{equation}
  \label{appexp8}
  {\cal E}^\prime(x) \simeq 1 - Kr_0\left[
    \frac{1}{2}\sqrt{\frac{\pi}{2}} x^{5/2} 
    e^{\displaystyle -x} + 36\pi g_1^2 \frac{1}{x^6}\right]
\end{equation}
Let us define $\tilde{x}$ such that for $x<\tilde{x}$ the first term in $\cal
E^\prime$ is larger than the second.  In this case the results of
section \ref{sec:inversedecays} hold.  We are interested now in the
situation in which $x_F > \tilde{x}$, so that it is $2\leftrightarrow 3$
scatterings that determine $Y_F(\infty)$.

The value of $\tilde{x}$ is given by the equation,
\begin{equation}
  \label{eq:x0}
\frac{1}{2}\sqrt{\frac{\pi}{2}} \tilde{x}^{5/2} e^{\displaystyle -\tilde{x}} = 36
\pi g_1^2 
  \frac{1}{\tilde{x}^6}~~. 
\end{equation}
Thus, we have,
\begin{equation}
  \label{x0final}
  \tilde{x} (g_1)\simeq 27.7 + \ell
  n\left(\frac{0.01}{g_1^2}\right)~~~(g_1\lesssim 0.1)~~.
\end{equation}

The minimum of $\cal E$ will then be at
\begin{equation}
  \label{xf8}
  x_F \simeq (36\pi Kr_0 g_1^2)^\frac{1}{6}~~,
\end{equation}
as long as $Kr_0$ is large enough so that the consistency condition
$x_F > \tilde{x}$ is satisfied.  For $g_1 = 0.1$, we obtain
\begin{equation}
  \label{eq:consis}
x_F > \tilde{x} ~~\Leftrightarrow~~  Kr_0 \gsim 4\times 10^{8}~~.
\end{equation}
At $x=x_F$ we have,
\begin{equation}
  \label{curv8}
  {\cal E}^{\prime\prime}(x_F) \simeq \frac{6}{x_F}~~.
\end{equation}
The minimum of ${\cal E}$ is therefore broad, and the steepest-descent
approximation cannot be applied.  However, in view of the preceeding
considerations, we expect $Y_F(\infty)$ to be exponentially damped in
this regime,
\begin{equation}
  \label{exp8}
  Y_F(\infty) \sim e^{-(36\pi Kr_0g_1^2)^{1/6}}
\end{equation}

As another example of $2\leftrightarrow 3$ scattering we consider the
diagram in figure \ref{fig:two2three}b.  
The last term in (\ref{eq:terms3+2+8}) must now be
substituted by,
\begin{equation}
  \label{term7}
  \mbox{}-\frac{2}{\pi^2} g_1^2 Kr_0 f(x) Y_F(x)~,
\end{equation}
where,
\renewcommand{\arraystretch}{2}
\begin{equation}
  \label{eq:resonance}
  f(x)\simeq \left\{
    \begin{array}{ll}
      \displaystyle \frac{8}{x^2} + 2 + O(x^2) & x \ll 1 \\
      \displaystyle \frac{2^{10} 3^2}{x^6} + O(x^{-8}) & x \gg 1
    \end{array}
  \right. ~~.  
\end{equation}
\renewcommand{\arraystretch}{1}
As in the previous case, for small $x$ this term can be neglected
compared to the point-scattering term on the second line of
(\ref{eq:terms3+2+8}).  For $x\sim 1$, cross-section
(\ref{eq:resonance}) is very suppressed due to the subtraction of
real-intermediate-particle contributions.  Only when $x\gsim 10$ can
this term be relevant.

The analysis follows the same lines as for the previous diagram, since
the asymptotic dependence is $x^{-6}$ in both cases.  This is, in
fact, a general result;  since the limit $x\gg 1$ corresponds to low
temperatures, {\em i.e.\/} low initial energies, the cross-section in
this limit is essentially determined by the final three-body
phase-space.  

The result in this case is,
\begin{equation}
  \label{eq:resx0}
  x_F \simeq 4 (Kr_0g_1^2)^{1/6}~~. 
\end{equation}
valid for $Kr_0 \gsim 10^7$.  Thus, we expect this $2\leftrightarrow
3$ process to start being relevant before the previous one as $Kr_0$
grows, and to have a somewhat stronger damping effect.

\begin{table}[htbp]
  \begin{center}
    \leavevmode
    \begin{tabular}{|r|r|r|r|r|}\cline{4-5}
       \multicolumn{3}{c}{ } & \multicolumn{2}{|c|}{$Y_F(\infty)$}\\ 
                                                        \hline
      \multicolumn{1}{|c}{$K$} & \multicolumn{1}{|c}{$r_0$} &
      \multicolumn{1}{|c}{$x_0$} & \multicolumn{1}{|c}{decays} & 
       \multicolumn{1}{|c|}{annihilations}\\\hline
       0.1 & $10^{-4}$ & 0.1 & $5.0\times10^{-8}$ & $3.9\times10^{-10}$
                                                                \\\hline 
       10 & $10^{-4}$ & 0.1 & $4.8\times10^{-8}$ & $4.5\times10^{-10}$
                                                                \\\hline
       100 & $10^{-4}$ & 0.1 & $3.5\times10^{-8}$ & $1.9\times10^{-10}$
                                                                 \\\hline
       100 & $5\times10^{-6}$ & 0.03 & $2.4\times10^{-9}$ &
                                                $4.9\times10^{-11}$\\\hline
       1000 & $10^{-4}$ & 0.1 & $1.4\times10^{-8}$ & $8.0\times10^{-12}$ 
                                                                  \\\hline
       1000 & $5\times10^{-6}$ & 0.03 & $2.0\times10^{-9}$ &
                                                  $1.9\times10^{-11}$\\\hline
    \end{tabular}
    \caption{Numerically obtained values of $Y_F(\infty)$ for both,
      $\chi$ decays/inverse decays and $\chi\bar\chi$ annihilations as
      source terms.  Fixed parameters are specified in 
      (\protect\ref{eq:arbitrary}).}
    \label{tab:vlues}
  \end{center}
\end{table}

\subsection{Annihilations and their back processes}
\label{sec:annbackpro}

The analysis of the term on the 4$^{\rm th}$ line of (\ref{eq:yf}),
due to $\chi\bar\chi$ annihilations and their back processes, as a
source of $F$-number generation can be carried out along lines
similar to those of the previous sections.  We shall only quote
numerical results here.  As expected on physical grounds, and can be
seen from table \ref{tab:vlues}, this term is quantitatively much less
important than decays and inverse decays.

\begin{figure}[htpb]
  \begin{center}
    \leavevmode
    \begin{picture}(100,75)(150,90)
      \put(10,0){\epsfbox{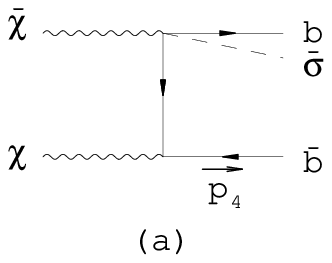}}
      \put(140,0){\epsfbox{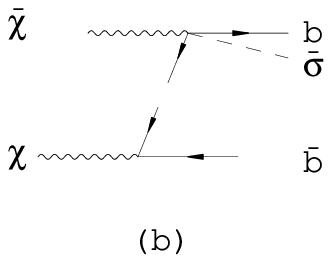}}
    \end{picture}
    \caption{(a) 
        $\chi\bar\chi$ annihilation diagram. (b) $\chi$ decay
        followed by point scattering.}
    \protect\label{fig:annihilations}
  \end{center}
\end{figure}

A Feynman diagram for $\chi\bar\chi$ annihilation is shown in figure
\ref{fig:annihilations}(a).  The intermediate fermion
can be on mass-shell, which makes the tree-level
cross-section singular.  Taking $p_4^0$ as a variable (see fig.\
\ref{fig:annihilations}(a)), at fixed center-of-mass energy $\sqrt{s}$,
the kinematical region where the intermediate particle can be on-shell
is given by
\begin{equation}
  \label{eq:kinter}
  \frac{\sqrt{s}-\sqrt{s-4 M}}{4} \leq p_4^0 \leq
  \frac{\sqrt{s}+\sqrt{s-4 M}}{4}~~.
\end{equation}
Notice that this interval is completely contained within the kinematical
domain $0\leq p_4^0\leq\sqrt{s}/2$.

The singularity of the amplitude at tree level corresponds to a
space-time-ordered sequence of $\chi$ decay and point scattering,
illustrated in figure \ref{fig:annihilations}(b), in which the
intermediate 
particle propagates over macroscopic distances.  The fact that
singularities in the physical region represent ordered sequences of
processes has been proved in general in \cite{colnor}.  Clearly, the
decay width of $\chi$ has to be included in the propagator in order to
obtain a finite cross-section.

\begin{figure}[p]
  \begin{center}
    \leavevmode
    \begin{picture}(100,75)(100,90)
      \put(0,0){\epsfbox{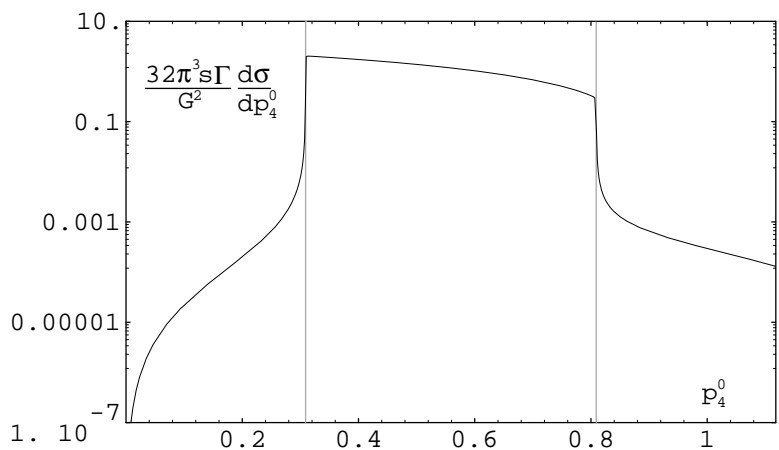}}
    \end{picture}
    \vspace{20pt}
    \caption{$\chi\bar\chi$ annihilation differential cross section
      for $s=5$, 
      $\Gamma_\chi=0.0004$ and $M=1$ ($G\equiv g_1 g_2/\Lambda$). The
      vertical lines show the 
      boundaries of the 
      kinematical domain of real-intermediate-particle exchange.}
    \label{fig:cross1}
  \end{center}
\end{figure}

\begin{figure}[p]
  \begin{center}
    \leavevmode
    \begin{picture}(100,75)(100,90)
      \put(0,0){\epsfbox{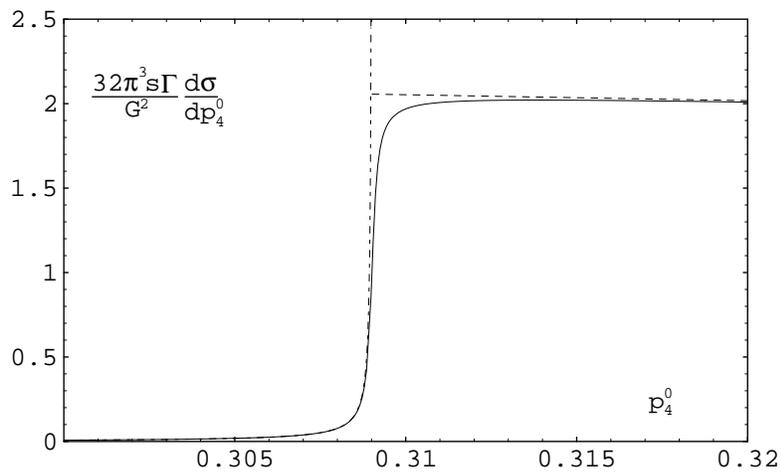}}
    \end{picture}
    \vspace{20pt}
    \caption{Solid line: same as previous figure. Dashed line:
      contribution of real particle exchange only, computed at
      $\Gamma_\chi=0$.  Dot-dashed line: contribution of virtual particle
      exchange only, at $\Gamma_\chi=0$.}
    \label{fig:cross2}
  \end{center}
\end{figure}

The differential cross-section $d\sigma/dp_4^0$ for $\chi\bar\chi$
annihilation is shown in figure \ref{fig:cross1}, for
 $\Gamma_\chi= 4\times 10^{-4} M$ which corresponds to the values of
parameters given in (\ref{eq:arbitrary}).  It is also shown in figure
\ref{fig:cross2} together with its value at zero width, and the
differential cross-section for exchange of a real intermediate
particle, also computed at $\Gamma_\chi=0$.  The former is singular at the
boundaries of interval (\ref{eq:kinter}), and the latter is zero
outside that interval.

In order to estimate the $F$-number generated by this process, we
set the subtracted differential cross-section
 $d\sigma^\prime/dp_4^0$ ({\em i.e.,} the virtual-particle exchange
differential cross-section) to be equal to $d\sigma/dp_4^0$, with the
above-mentioned value for $\Gamma_\chi$, outside the 
kinematical limits for real-intermediate-particle exchange and zero
inside that interval. 

The evolution equation for $Y_F$, with
 $\chi\bar\chi$ annihilations as the only source term, can then be
 numerically solved with the results 
shown in table \ref{tab:vlues}.  This gives an estimate of the
importance of this term relative to the other source term. The $F$-number
generated in this case 
is negligible compared to that arising from decays and inverse decays,
except for very large values of $K$ and very small values of $r_0$.  In
this last case both the decays- and annihilations-generated
 $F$-numbers are small themselves, since the system never departs much
 from equilibrium.

\section{Asymmetric initial conditions.}
\setcounter{equation}{0}
\label{sec:asymm}

We consider now the case in which the initial value for the density
$Y_F$ is different from zero.  To be concrete, we take $Y_{F0}>0$.  It
is clear from the foregoing analysis that the processes which will be
relevant to the erasure of $Y_{F0}$ are point scatterings.  Therefore,
we have, 
\begin{equation}
  \label{eq:asymyf}
  Y_F(\infty) = Y_{F0} \exp\left\{-192 K r_0
    \int_{x_0}^{\infty}\!\!dz~\frac{1}{z^2} \left(\rule{0ex}{2ex}
      Y_+(z) + 
      Y_+^{eq}(z)\right)\right\} + Y_F^{\rm sym}(\infty),
\end{equation}
where $Y_F^{\rm sym}(\infty)$ refers to the value of $Y_F(\infty)$
obtained in the previous sections with $Y_{F0}=0$.

We can easily evaluate the exponent as,
\begin{equation}
  \label{eq:safeapp}
  \int_{x_0}^{\infty}\!\!dz~\frac{1}{z^2} \left( Y_+(z) +
      Y_+^{eq}(z)\right)\simeq \frac{1}{x_0}
\end{equation}
to write,
\begin{equation}
  \label{asymyfinal}
  Y_F(\infty)\simeq Y_{F0} \exp\left(-\frac{192 K r_0}{x_0}\right) +
  Y_F^{\rm sym}(\infty).
\end{equation}

As mentioned in section \ref{sec:paramandincond}, the value of $x_0$
cannot be chosen arbitrarily small, since lagrangian
(\ref{eq:lagrangian}) is not applicable at energies higher than 
${\cal O}(\Lambda)$.  More precisely, tree-level diagrams such as
non-renormalizable decays and point scatterings violate the unitarity
bound at an energy scale of order \mbox{$\sim\Lambda/g_2$}.  The appearance
of $g_2$ here should not be surprising, since $\Lambda$ enters 
 $\cal L$ only in the combination $g_2/\Lambda$.  

\begin{figure}[tpbh]
  \begin{center}
    \leavevmode
    \begin{picture}(300,175)(50,62.5)
      \put(50,0){\epsfbox{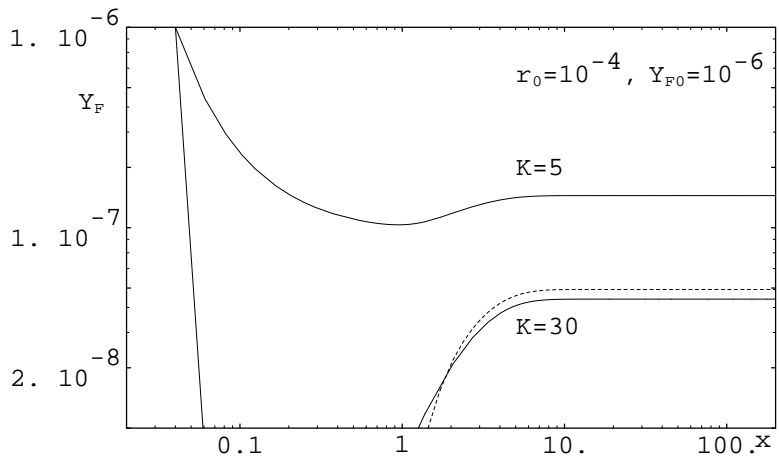}}
    \end{picture}
    \parbox{300pt}{\protect\caption{Evolution of $Y_F$ from a non-zero initial
        value (solid lines).  The dashed line shows $Y_F(x)$ for $K=5$
        in the symmetric case $Y_{F0}=0$. $x_0=0.04$. }
   \protect\label{fig:asym1}}
  \end{center}
\end{figure}

Thus, the minimal value of $x_0$ we can choose is, 
\begin{equation}
  \label{eq:minx0}
  x_0 \sim\frac{g_2 M}{\Lambda}\simeq 30 g_1 \sqrt{r_0},
\end{equation}
where we used (\ref{eq:x0compelling}) in the last approximate
equality.  With this value of $x_0$ we have, finally,
\begin{equation}
  \label{eq:yfinftyfinally}
  Y_F(\infty)\simeq Y_{F0} \exp\left(\frac{-4 K\sqrt{r_0}}{g_1}\right) +
    Y_F^{\rm sym}(\infty).
\end{equation}
Depending on the values of the parameters $K$, $r_0$, $g_1$ and $\eta$
(since $Y_F^{\rm sym}(\infty)\propto\eta$), and of $Y_{F0}$, one of the
two terms will dominate.  Only when the first one is much smaller than
the second will this model display dynamical,
initial-condition-independent generation of global charge $Y_F$.  This
is illustrated in figure \ref{fig:asym1}, for $K=30$ and 5.
In the first case the initial condition is completely erased, and only
partially in the second.

\section{Final Remarks}
\setcounter{equation}{0}

We considered a schematic model for dynamical generation of a global
charge.  The interactions giving rise to the violation of $F$-number
are given by an operator of dimension 5, whereas the renormalizable
piece of the Lagrangian is $F$-symmetric.  Our model is a special case
of the standard out-of-equilibrium decays scenario
\cite{kotu,kotu2,oliv}, with the particularity that the heavy $\chi$
boson can decay through two channels only one of which, the one
mediated by the effective interactions, violates conservation of
$F$-number.

We introduced $CP$ violation through a parametrization of the matrix
elements for $\chi$ decays and $\chi\bar\chi$ annihilations consistent
with unitarity and $CPT$ (to the order considered here in the coupling
constants) \cite{kowo,frol}.  We have, then, two independent $CP$
violation parameters, $\eta$ and $\xi$, which represent the net
$F$-number generated in decay and annihilation reactions,
respectively. All other possible violations of $CP$ are related to
$\eta$ and $\xi$ through unitarity and $CPT$.  We made one additional
assumption, that $\xi$ is constant.  Genuine $CP$ violation would
require explicit modelling \cite{kotu,kotu2}, with the introduction of
more fields and coupling constants.  We preferred not to do so, to
keep our model as simple as possible.

The model assumes the standard Friedmann-Robertson-Walker cosmology
\cite{kotu,kotu2,oliv,frol}, which enters the evolution equations
explicitly through the Hubble parameter $H$.  Furthermore, other
species and interactions are assumed to exist, which do not change
$F$-number, give a $g_{\star}\sim {\cal O}(100)$, and maintain light
degrees of freedom in local thermal equilibrium.  In order to
establish the evolution equations and make them tractable we made two
approximations that are known to be valid in more realistic models
such as GUT-based baryogenesis models \cite{kowo,frol,hako}.  These
are the neglect of degeneracy factors and the use of Maxwell-Boltzmann
equilibrium distributions.  We also neglected processes of order
higher than the first in $g_1$ and $g_2$.  In view of the
negligibly small effects of $2\rightarrow3$ scatterings for all
reasonable values of parameters, this approximation is justified.

The model has five parameters, which can be chosen as explained in
section \ref{sec:paramandincond}.  We kept $g_1$, $\eta$ and $\xi$
fixed (see \ref{eq:arbitrary}), and studied the final value of the
global charge for broad ranges of values of $K$ and $r_0$.  Since we
set the coupling constant for non-renormalizable interactions $g_2 =
0.1 = g_{1}$, the condition $M < \Lambda$ leads to an upper bound for
$r_0$, $r_0 \lesssim 5\times 10^{-4}$.

Besides those five parameters, initial conditions $x_0$, $Y_{F0}$ must
be chosen.  In the symmetric case $Y_{F0} \ll Y_F(\infty)$ the final
value of $Y_F(\infty)$ is insensitive to the previous value of $x_0$.
We have used $x_0\sim M/\Lambda$ in this case.  For asymmetric initial
conditions the choice of $x_0$ is more delicate.  Whether large
initial asymmetries are completely erased or not depends on the value
of $Y_{F0}$, $K$ and $r_0$, but also on $x_0$.  Due to the nature of
our model, we cannot approximate $x_0\simeq 0$ as in other simple
models \cite{kotu}, since some of the terms in the evolution equation
(notably point scatterings) are singular at $x=0$.  Thus, we had to
determine the minimal value $x_0$ can take.  This is the value at
which non-renormalizable processes such as point-scatterings start
violating unitarity bounds, $x_0\sim g_2 M/\Lambda$.  For $x$ much
less than this value, the description of $F$ violating interactions as
effective operators ceases to be valid.

The case of symmetric initial conditions is the more interesting,
since it displays purely dynamical generation of global charge.  The
appearance of an additional parameter $r_0$ in our model, besides $K$,
which determines the effectiveness of damping, introduces some
differences with respect to renormalizable models.  As shown in
section \ref{sec:yfevolsym}, damping is effective only when $Kr_0\gsim
10^{-2}$, so even for relatively large values of $K$ we can have
undamped generation of $F$-charge.  The largest $Y_F(\infty)$
attainable in this case is $\eta r_0$.  

Also as a result of having a branching ratio $r_0$, exponential
damping due to $2\rightarrow 3$ scatterings is effectively postponed
to very large values of $K$, since $Kr_0$ itself must be large for
these processes to be in thermal equilibrium.  We mentioned in section
\ref{sec:yfevolsym} that the number of $2\rightarrow 3$ scattering
channels is rather large in our model, and considered two specific
examples which showed that these processes are irrelevant for
$K\lesssim 10^9$. (Notice, however, that they are important in
deriving the Boltzmann equations, as explained in section
\ref{sec:boleq} \cite{kowo}).

An important class of processes not present in renormalizable models
is two-body point scatterings.  (These have been considered in the
framework of SUSY GUTs in \cite{frtu,kora}.)  These scatterings turn
out to be the most efficient damping process at low $x$.  Thus, they
are crucial for erasing initial asymmetries.  For symmetric initial
conditions, $2 \rightarrow 2$ scatterings give algebraic damping
($\sim 1/K$) for $Kr_0 \gsim 10^{-2}$.  Inverse decays are also
important for $Kr_0 \gsim 7$.  They are in equilibrium for $x\sim 1$,
for large enough $Kr_0$, and also produce $1/K$ damping of the
generated $Y_F$.

Besides $\chi$ decays and inverse decays, $\chi\bar\chi$ annihilations
and their back processes are also a source of $F$-generation.  Since
$CP$ violation in this case is independent of that in decays, it is
always possible to choose parameters in such a way that annihilations
be the dominant source term (by setting $\xi\gg\eta$).  In the absence
of explicit modelling of $CP$ violation this situation cannot be ruled
out.  We considered this possibility artificial, and used
$\xi\sim\eta$ in our computations in order to be able to compare the
relative importance of both source terms.  As shown in section
\ref{sec:annbackpro}, the annihilations term is negligible compared to
decays over a wide range of values of $K$ and $r_0$.  When $K$ is
large and $r_0$ small enough to block damping ($K r_0\lesssim
10^{-2}$), the $F$-number generated by annihilations is comparable to
that generated by decays, typically one order of magnitude smaller.

\appendix

\section{Evolution equations}
\label{sec:appa}
\setcounter{equation}{0}

   Here, we give the full evolution equations of $n_\chi, n_\sigma, 
n_b$ and show that the $F$-number asymmetry only depends on $Y_\sigma 
- Y_{\bar\sigma}$.
We define
\bea
\Theta(\alpha_1\alpha_2\cdots \leftrightarrow 
\beta_1\beta_2\cdots)
&\equiv &
\int \Da \DA \cdots \Dc \DC \cdots 
\nonumber\\
&& \times \left [ f_{\alpha_{1}} f_{\alpha_{2}} \cdots
|M (\alpha_1\alpha_2\cdots \rightarrow \beta_1\beta_2\cdots)|^2 
\right. \nonumber\\
&& \left. - f_{\beta_{1}} f_{\beta_{2}} \cdots 
|M (\beta_1\beta_2\cdots \rightarrow \alpha_1\alpha_2\cdots)|^2 \right ]
\nonumber\\
& = & - \Theta(\beta_1\beta_2\cdots\leftrightarrow 
\alpha_1\alpha_2\cdots),
\eea
where $\alpha_1\alpha_2\cdots$, $\beta_1\beta_2\cdots$ label different particles, and $\Theta'$ is given by a similar expression where $M'$
replaces $M$ (namely the amplitude with on-mass shell propagators substracted)
. The full $n_\chi$ evolution equations is,
\bea
{dn_\chi\over dt} + Hn_\chi
& = & - \Theta_{D+ID}(\chi \leftrightarrow b d) 
- \Theta_{NRD+NRID}(\chi \leftrightarrow b d \sigma) 
\nonumber\\
&& - \Theta_{PS}(\bar\sigma \chi \leftrightarrow b d)
- 2\Theta_{PS}(\bar b \chi \leftrightarrow d \sigma)
- 2\Theta_{AN}(\chi \bar\chi \leftrightarrow b \bar b) 
\nonumber\\
&&- 2\Theta'_{NRAN+NRCC}(\chi \bar\chi \leftrightarrow b \bar b \sigma) 
- 2\Theta'_{NRAN+NRCC}(\chi \bar\chi \leftrightarrow b \bar b \bar\sigma) \nonumber\\
&&- 2\Theta_{NRCC}(\sigma \chi \bar\chi \leftrightarrow b \bar b)
- 2\Theta_{NRCC}(\bar\sigma \chi \bar\chi \leftrightarrow b \bar b) 
\nonumber\\
&&- 2\Theta_{NRCC}(b \chi \bar\chi \leftrightarrow b \sigma)
- 2\Theta_{NRCC}(b \chi \bar\chi \leftrightarrow b \bar\sigma) 
\nonumber\\
&&- 2\Theta_{NRCC}(\bar b \chi \bar\chi \leftrightarrow \bar b \sigma)
- 2\Theta_{NRCC}(\bar b \chi \bar\chi \leftrightarrow \bar b \bar\sigma),
\eea
where the factor of 2 in $PS$ and $AN$ ($R$ or $NR$) terms accounts
for similar processes with $b$ and $d$ exchanged. The $n_{\bar\chi}$
evolution equation is obtained by replacing in (A.2) all particles by
their antiparticles.

The full $n_b$ evolution equation is,
\bea
{dn_b\over dt} + Hn_b
& = & + \Theta_{D+ID}(\chi \leftrightarrow b d)
+ \Theta_{NRD+NRID}(\chi \leftrightarrow b d \sigma)
+ \Theta_{AN}(\chi \bar\chi \leftrightarrow b \bar b) 
\nonumber\\ 
&&+ \Theta_{PS}(\bar d \chi \leftrightarrow b \sigma)
- \Theta_{PS}(b \bar\chi \leftrightarrow d \sigma)
+ \Theta_{PS}(\bar\sigma \chi \leftrightarrow b d)
\nonumber\\
&&+ \Theta_{NRAN+NRCC}(\chi \bar\chi \leftrightarrow b \bar b \sigma)
+ \Theta_{NRAN+NRCC}(\chi \bar\chi \leftrightarrow b \bar b \bar\sigma) \nonumber\\
&&- \Theta_{NRCC}(b \bar b \leftrightarrow \sigma \chi \bar\chi)
- \Theta_{NRCC}(b \bar b \leftrightarrow \bar\sigma \chi \bar\chi) 
\nonumber\\
&&+ \Theta_{NRCC}(\sigma \chi \leftrightarrow b \bar b \chi)
+ \Theta_{NRCC}(\sigma \bar\chi \leftrightarrow b \bar b \bar\chi) 
\nonumber\\
&&+ \Theta_{NRCC}(\bar\sigma \chi \leftrightarrow b \bar b \chi)
+ \Theta_{NRCC}(\bar\sigma \bar\chi \leftrightarrow b \bar b \bar\chi) \nonumber\\
&&- \Theta_{NRS}(b \bar b \leftrightarrow d \bar d \sigma)
- \Theta_{NRS}(b \bar b \leftrightarrow d \bar d \bar\sigma)
+ \Theta_{NRS}(d \bar d \leftrightarrow b \bar b \sigma) 
\nonumber\\
&&+ \Theta_{NRS}(d \bar d \leftrightarrow b \bar b \sigma)
+ \Theta_{NRS}(d \sigma \leftrightarrow b \bar b d)
+ \Theta_{NRS}(d \bar\sigma \leftrightarrow b \bar b d) 
\nonumber\\
&&+ \Theta_{NRS}(\bar d \sigma \leftrightarrow b \bar b \bar d)
+\Theta_{NRS}(\bar d\bar\sigma\leftrightarrow b\bar b\bar d).
\eea
Similar expressions hold for $n_d$ (switching $b$ and $d$), and for
$n_{\bar b}$ , $n_{\bar d}$ (replacing the particles by their
antiparticles.  Using (A.2) and (A.3) one can show $n_b-n_{\bar b}$
and $n_\chi-n_{\bar\chi}$, evolve equally, i.e.
\bea
{d(n_b-n_{\bar b})\over dt} + H(n_b-n_{\bar b})
& = &{d(n_d-n_{\bar d})\over dt} + H(n_d-n_{\bar d}) 
\nonumber\\
& = & + \Theta_{D+ID}(\chi \leftrightarrow b d)
- \Theta_{D+ID}(\bar\chi \leftrightarrow \bar b \bar d) 
\nonumber\\
&&+ \Theta_{NRD+NRID}(\chi \leftrightarrow b d \sigma)
 - \Theta_{NRD+NRID}(\bar\chi \leftrightarrow \bar d \bar d \bar\sigma)
\nonumber\\
&&+ \Theta_{PS}(\bar\sigma \chi \leftrightarrow b d)
- \Theta_{PS}(\bar\sigma \chi \leftrightarrow \bar b \bar d)
\nonumber\\
&&+ 2\Theta_{PS}(\bar b \chi \leftrightarrow d \sigma)
- 2\Theta_{PS}(b \bar\chi \leftrightarrow \bar d \sigma)
\nonumber\\
& = & - \left [ {d\over dt} (n_\chi-n_{\bar\chi}) +
H(n_\chi-n_{\bar\chi}) \right ].
\eea
Hence
\be
{d(Y_b-{Y_{\bar b}})\over dt} = -
{d(Y_\chi-{Y_{\bar\chi}})\over dt} =
{d(Y_d -{Y_{\bar d}})\over dt}, 
\ee
and since $Y_F = 4(Y_{\bar\sigma} - Y_\sigma) + 2 (Y_\chi-Y_{\bar\chi}) + (Y_b-Y_{\bar b}) + (Y_d-Y_{\bar d})$,
$Y_F$ is given by $Y_{\bar\sigma} - Y_\sigma$.

The full $n_\sigma$ evolution equation is
\bea
{dn_\sigma\over dt} + {Hn_\sigma} 
& = & + \Theta_{NRD+NRID}(\chi \leftrightarrow b d \sigma)
+ 2\Theta_{PS}(\bar b \chi \leftrightarrow d \sigma)
+ \Theta_{PS}(\bar b \bar d \leftrightarrow \sigma \bar\chi)
\nonumber\\
&& + 2\Theta'_{NRAN+NRCC}(\chi \bar\chi \leftrightarrow b \bar b \sigma)
+ 2\Theta_{NRCC}(b \chi \leftrightarrow b \sigma \chi)
\nonumber\\
&&+ 2\Theta_{NRCC}(b \bar\chi \leftrightarrow b \sigma \bar\chi)
+ 2\Theta_{NRCC}(\bar b \chi \leftrightarrow \bar b \sigma \chi)
\nonumber\\
&&+ 2\Theta_{NRCC}(\bar b \bar\chi \leftrightarrow \bar b \sigma \bar\chi)
- \Theta_{NRCC}(\sigma \chi \leftrightarrow b \bar b \chi)
\nonumber\\
&& - 2\Theta_{NRCC}(\sigma \bar\chi \leftrightarrow b \bar b \bar\chi)
+ 2\Theta_{NRCC}(b \bar b \leftrightarrow \sigma \chi \bar\chi) 
\nonumber\\
&&- 2\Theta_{NRCC}(b \sigma \leftrightarrow b \chi \bar\chi)
- 2\Theta_{NRCC}(\bar b \sigma \leftrightarrow \bar b \chi \bar\chi)
\nonumber\\
&&+ \Theta'_{NRS}(b d \leftrightarrow d b \sigma)
+ \Theta'_{NRS}(\bar b \bar d \leftrightarrow \bar b \bar d \sigma)
\nonumber\\
&& + 2\Theta_{NRS}(b \bar d \leftrightarrow  b \bar d \sigma)
+ 2\Theta_{NRS}(b \bar b \leftrightarrow d \bar d \sigma)
\nonumber\\
&&- 2\Theta_{NRS}(b \sigma \leftrightarrow b d \bar d)
- 2\Theta_{NRS}(\bar b \sigma \leftrightarrow \bar b d \bar d).
\eea
Changing all particles by their antiparticles in (A.6) one obtains the equation for $n_{\bar\sigma}$.

The evolution equation of $Y_-$ is,
\bea
dY_-\over dx
& = & -{\langle\Gamma_\chi \rangle\over xH} \left [Y_-(1+2Y_+^{eq}) +r_0 Y_F Y_+^{eq}(1-2r_0 Y_-) \right ]
\nonumber\\
&& -24r_0 x^{-3} {\langle \Gamma_\chi \rangle\over H} \left [Y_-(1+2Y_+^{eq}) +Y_F Y_+ \rangle \right]
\nonumber\\
&& -96r_0 x^{-4} {\langle \Gamma_\chi \rangle\over H} \left [Y_-(2+Y_++2Y_+^{eq}) +Y_F(Y_++Y_+^{eq}) \rangle \right].
\eea

\section{ $CP$ violation}
\label{sec:appb}
\setcounter{equation}{0}

All the processes we have considered in this paper are upto the order
$(g_2/\Lambda)^2$.
In this Appendix, we will use unitarity to check the possibility of $CP$ 
violating differences $[|M(a\rightarrow b)|^2 - |M(b\rightarrow a)|^2]$ 
upto the order $(g_2/\Lambda)^4$.

In order to have $CP$ violation in a process $|a\rangle 
\rightarrow|b\rangle$, there must be other final states besides 
$|b\rangle$ for $|a\rangle$ to go to and other initial state besides 
$|a\rangle$ that can go to $|b\rangle$.
Namely, unitarity imposes the following relations,
\be
\sum_j|M(a\rightarrow j)|^2 =
\sum_j|M(\bar a\rightarrow \bar j)|^2
\ee
\be
\sum_i|M(i\rightarrow b)|^2 = \sum_i|M(\bar i \rightarrow \bar b)|^2
\ee
where $\bar a, \bar b, \bar i, \bar j$ denote the $CP$ conjugates of $a, b, 
i, j$ respectively.
Therefore, if $|b\rangle$ is the unique state of $|a\rangle$ or 
$|a\rangle$ the unique initial state for $|b\rangle$, then either (B.1) 
or (B.2) imply
\[
|M(a\rightarrow b)|^2 = |M(\bar a\rightarrow \bar b)|^2~~, 
\]
which is equivalent to the requirement of $CP$ conservation in the 
process.
Using this condition, we find most processes do not violate $CP$ upto the 
order of $(g_2/\Lambda)^4$.

Let us look at some examples:

  1. Consider $|\bar b \chi \rangle$ as initial state.
By (B.1), we get
\bea
&&\int \left [ \Dd \Ds |M(\bar b \chi \rightarrow d \sigma)|^2
+ \DB \Ds \Dx
|M(\bar b \chi \rightarrow \bar b \sigma \chi)|^2
\right. \nonumber\\
&& \left. \hspace{1.1in} + \DB \DS \Dx
|M(\bar b \chi \rightarrow \bar b \bar\sigma \chi)|^2 \right ]
\nonumber\\
&=&\int \left [ \DD \DS
|M(b \bar\chi \rightarrow \bar d \bar\sigma)|^2
+ \Db \DS \DX
|M(b \bar\chi \rightarrow b \bar\sigma \bar\chi)|^2
\right. \nonumber\\
&& \left. \hspace{1.1in} + \Db \Ds \DX 
|M(b \bar\chi \rightarrow b \sigma \bar\chi)|^2 \right ]
\eea
that may seem to allow $CP$ violation.
But with $|\bar b \sigma \chi \rangle$ and $|\bar b \bar\sigma \chi \rangle$ as final states by (B.2) we get
\bea
|M(\bar b \chi \rightarrow \bar b \sigma \chi)|^2 
& = & |M(b \bar\chi \rightarrow b \bar\sigma \bar\chi)|^2 ,\\
|M(\bar b \chi \rightarrow \bar b \bar\sigma \chi)|^2 
& = & |M(b \bar\chi \rightarrow b \sigma \bar\chi)|^2
\eea
and using (B.3), we have
\be
|M(\bar b \chi \rightarrow d \sigma)|^2 = |M(b \bar\chi \rightarrow \bar d \bar\sigma)|^2
\ee
Therefore there is no $CP$ violation in any of the three processes in 
(B.3).

2.  With the assumption of no asymmetry of $b$ and $d$ in all 
processes upto the order of $(g_2/\Lambda)^2$, i.e.
\be
|M(\bar\sigma \chi \rightarrow b \bar b \chi)|^2 =
|M(\bar\sigma \chi \rightarrow d \bar d \chi)|^2
\ee
we find again no $CP$ violation in processes that involve the state
$|\bar\sigma X\rangle$.
In this case, the relation (B.1) is
\bea
&&\int \left [ \Db \Dd 
|M(\bar\sigma \chi \rightarrow b d)|^2
+ \Db \DB \Dx 
|M(\bar\sigma \chi \rightarrow b \bar b \chi)|^2 
\right. \nonumber\\
&& \left. \hspace{1in} + \Dd \DD \Dx
|M(\bar\sigma \chi \rightarrow d \bar d \chi)|^2 \right ] 
\nonumber\\
& = & \int \left [ \DB \DD 
|M(\sigma \bar\chi \rightarrow \bar b \bar d)|^2 
+ \Db \DB \DX 
|M(\sigma \bar\chi \rightarrow b \bar b \bar\chi)|^2 
\right. \nonumber\\
&& \left. \hspace{1in} + \Dd \DD \DX
|M(\sigma \bar\chi \rightarrow d \bar d \bar\chi)|^2 \right ] .
\eea
Now, consider $|b\bar b \chi\rangle$ as the final state then (B.2) gives
\be
|M(\sigma \chi \rightarrow b \bar b \chi)|^2
= |M(\bar\sigma \chi \rightarrow b \bar b \chi)|^2
= |M(\sigma \bar\chi \rightarrow b \bar b \bar\chi)|^2
\ee
and with $|\sigma\chi\rangle$ as initial state, (B.1) and (B.7) yield
\be
|M(\sigma \chi \rightarrow b \bar b \chi)|^2
= |M(\sigma \chi \rightarrow d \bar d \chi)|^2
= |M(\bar\sigma \bar\chi \rightarrow b \bar b \bar\chi)|^2
= |M(\bar\sigma \bar\chi \rightarrow d \bar d \bar\chi)|^2
\ee
Then, combining (B.9) and (B.10), one obtain
\be
|M(\bar\sigma \chi \rightarrow b \bar b \chi)|^2
= |M(\sigma\bar \chi \rightarrow b \bar b \bar\chi)|^2 ,
\ee
and (B.11) together with (B.8) give
\be
|M(\bar\sigma \chi \rightarrow b d)|^2
= |M(\sigma \bar\chi \rightarrow \bar b \bar d)|^2
\ee

To sum up, we find that only the processes 
$| \chi \rangle \rightarrow |b d \rangle$, 
$| \chi \rangle \rightarrow |b d \sigma \rangle$,
$|b d \rangle \rightarrow |b d \sigma \rangle$, 
$|b \bar b \rangle \rightarrow |d \bar d \sigma \rangle$, 
$|d \bar d \rangle \rightarrow |b \bar b \sigma \rangle$, 
$|\chi \bar\chi \rangle \rightarrow |d \bar d \sigma \rangle$,
$|\chi \bar\chi \rangle \rightarrow |b \bar b \sigma \rangle$, 
and their $CP$ conjugates violate $CP$.
We have related the $CP$ violation in $|\chi \rangle \rightarrow |b d
\rangle$ and $|\chi \rangle \rightarrow |b d \sigma \rangle$ and their
$CP$ conjugates by using the branching ratios $r$ and $\bar r$.  The
$CP$ violation in $|\chi \rangle \rightarrow |b d \sigma \rangle$ and
the violation in $|b d \rangle \rightarrow |b d \sigma \rangle$ are
related in (\ref{eq:decscatt}).
With $|b \bar\sigma \rangle$ as the final state in (B.2), we obtain
\bea
&&\int \left [ \Dx \DX
|M'(\chi \bar\chi \rightarrow b \bar b \sigma)|^2
+ \Dd \DD
|M(d \bar d \rightarrow b \bar b \sigma)|^2 \right ]
\nonumber\\
& = &
\int \left [ \Dx \DX
|M'(\chi \bar\chi \rightarrow b \bar b \bar\sigma)|^2
+ \Dd \DD
|M(d \bar d \rightarrow b \bar b \bar\sigma)|^2 \right ] ,
\eea
namely,
\bea
&&\int \Dx \DX
\left [ |M'(\chi \bar\chi \rightarrow b \bar b \bar\sigma)|^2
- |M'(\chi \bar\chi \rightarrow b \bar b \sigma)|^2 \right ]
\nonumber\\
& = &- \int \Dd \DD 
\left [ |M(d \bar d \rightarrow b \bar b \bar\sigma)|^2 
- |M(d \bar d \rightarrow b \bar b \sigma)|^2 \right ] 
\nonumber\\
& = &2 \xi \int \Dx \DX
|M(\chi \bar\chi \rightarrow b \bar b \sigma)|^2
\label{eq:B}
\eea
where $\xi$ is defined in (\ref{eq:xi}).
We have a similar equation for $|d \bar d \sigma \rangle$ as final
state.  Note that in (\ref{eq:B}) we take $|M(\chi \bar\chi \rightarrow
d \bar d \sigma)|^2 = |M(\chi \bar\chi \rightarrow b \bar b
\sigma)|^2$, since we assume no asymmetry of $b$ and $d$ in all the
processes upto order $(g_2/\Lambda)^2$.

\section{Approximate solution for $Y_{+}$}
\label{sec:appc}
\setcounter{equation}{0}

The equation for $\Delta(x)=Y_+-Y_+^{eq}$, neglecting terms proportional to
$Y_{B}$ and $Y_{-}$, is given by,
\begin{equation}
\label{eq:delta}
\frac{d\Delta}{dx} = - K g(x) \Delta(x) - \frac{dY_+^{eq}}{dx}~,
\end{equation}
where we used the notation $g(x)=x K_{1}(x)/K_{2}(x)$.  Defining
$\Delta_{K} (x) = K \Delta(x/K)$, we get the equation,
\begin{equation}
\frac{d\Delta_{K}(x)}{dx} = - g\left(\frac{x}{K}\right) \Delta_{K}(x) -
\frac{dY_+^{eq}}{dx}\left(\frac{x}{K}\right)
\end{equation}
with the initial condition $\Delta_{K}(Kx_{0}) = K\Delta(x_{0}) =0$.

For very large $K$, we can expand the coefficient functions $g(x/K)$
and $dY_+^{eq}(x/K)/dx$ in
this equations about $x/K=0$ (i.e. for  $K \gg x$), solve the resulting
equation for $\Delta_{K}$, and transform  back to $\Delta$ to obtain,
\bea
\Delta(x) & = & 1/8 \left( x^2 ~
\mbox{}_{1}F_{1}(\frac{2}{3},\frac{5}{3},\frac{Kx^{3}}{6})
 - x_{0}^{2}~
\mbox{}_{1}F_{1}(\frac{2}{3},\frac{5}{3},\frac{Kx_{o}^{3}}{6})\right)
\times\nonumber\\ 
  &  & \exp(-K x^3/6)
\eea
where ${}_{1}F_{1}$ is a confluent hypergeometric function \cite{abram}.

For $K x^3 \gg 1$ (i.e. $K^{-1/3} \ll x \ll K$) we obtain,
\begin{equation}
\Delta(x) \approx \frac{1}{2} \frac{1}{K x}
\end{equation}
which coincides with the result (see \cite{kotu}, eqn.\ (6.29)), 
\be
\Delta(x)=Y_+^{eq}(x)  /(K x)
\ee
if we approximate,
$Y_+^{eq}\approx 1/2$, valid for $x<1$. Notice that in our model
 $\chi$ is a scalar, so $Y_{+}^{eq}(x\ll 1)=1/2$, instead of 1. 
This suggests that we take,
for large $K$, 
\bea
\Delta(x) & =  & \frac{1}{4}  Y_+^{eq}(x)\left( x^2~ 
\mbox{}_{1}F_{1}(\frac{2}{3},\frac{5}{3},\frac{Kx^{3}}{6}) -
x_{0}^{2}~
\mbox{}_{1}F_{1}(\frac{2}{3},\frac{5}{3},\frac{Kx_{o}^{3}}{6})\right)
\times\nonumber\\ &  & \exp(-K x^3/6).
\eea

This expression is a very accurate representation of $\Delta$ for
large values of $K$, which already at $K=10$ departs by about 5\% from
the numerical solution of equation (\ref{eq:delta}) at $x\simeq1$, 
the region where the error is largest.  


\begin{thebibliography}{99}
\bibitem{geho} G.\ Gelmini and R.\ Holman, Phys.\ Lett.\ {\bf B316},
  (1993), 61.
\bibitem{kotu} E.\ Kolb and M.\ Turner, {\em The Early Universe},
  Addison-Wesley, New York, (1990).
\bibitem{kowo} E.\ Kolb and S.\ Wolfram, Nucl.\ Phys.\ {\bf B172},
  (1980), 224.
\bibitem{frol} J.\ Fry, K.\ Olive and M.\ Turner, Phys.\ Rev.\ {\bf
    D22}, (1980), 2977.
\bibitem{hako} J.\ Harvey {\em et.\ al.}, Nucl.\ Phys.\ {\bf B201},
  (1982), 16.
\bibitem{sakh} A.\ Sakharov, JETP Lett.\ {\bf 5}, (1967), 24.
\bibitem{wein} S.\ Weinberg, Phys.\ Rev.\ Lett.\ {\bf 42}, (1979),
  850, and references therein.
\bibitem{kotu2} E.\ Kolb and M.\ Turner, Ann.\ Rev.\ Nucl.\ Part.\
  Sci.\ {\bf 33}, (1983), 645.
\bibitem{oliv} K.\ Olive, {\em Big Bang Baryogenesis}, in Proceedings
  of the 33$^{\rm rd}$ International Winter School on Nuclear and Particle
  Physics, Schladming, Austria, H.\ Latal and W.\ Schweiger eds.,
  Springer-Verlag, Berlin, (1994).
\bibitem{frtu} J.\ Fry and H.\ Turner, Phys.\ Lett.\ {\bf B125},
  (1983), 379.
\bibitem{kora} E.\ Kolb and S.\ Raby, Phys.\ Rev.\ {\bf D27}, (1983),
  2990. 
\bibitem{abram} M.\ Abramowitz and I.\ Stegun, {\em Handbook of
    Mathematical Functions}, 9$^{\mbox{\small th}}$ edition, Dover, 
    New York, (1970).
\bibitem{clka} J.\ Cline, K.\ Kainulainen and K.\ Olive, Phys.\ Rev.\
  {\bf D49}, (1994), 6394. 
\bibitem{colnor} S.\ Coleman and R.\ Norton, Nuov.\ Cim.\ {\bf 38},
  (1965), 438.
\end{thebibliography}
\end{document}